\documentclass[10pt]{article}
\usepackage{bbm}
\usepackage[final]{graphics}
\usepackage{amsmath}
\usepackage{amsfonts,amsbsy}
\usepackage{amssymb}

\def\empile#1\over#2{\mathrel{\mathop{\kern 0pt#1}\limits_{#2}}}
\def\bs{\boldsymbol}

\newcommand{\slv}{\raise.15ex\hbox{$/$}\kern-.53em\hbox{$v$}}
\newcommand{\slF}{\raise.15ex\hbox{$/$}\kern-.53em\hbox{$F$}}
\newcommand{\slL}{\raise.15ex\hbox{$/$}\kern-.53em\hbox{$L$}}
\newcommand{\slP}{\raise.15ex\hbox{$/$}\kern-.53em\hbox{$P$}}
\newcommand{\slp}{\raise.15ex\hbox{$/$}\kern-.53em\hbox{$p$}}
\newcommand{\slq}{\raise.15ex\hbox{$/$}\kern-.53em\hbox{$q$}}
\newcommand{\slR}{\raise.15ex\hbox{$/$}\kern-.53em\hbox{$R$}}
\newcommand{\slQ}{\raise.15ex\hbox{$/$}\kern-.53em\hbox{$Q$}}
\newcommand{\slK}{\raise.15ex\hbox{$/$}\kern-.53em\hbox{$K$}}
\newcommand{\slk}{\raise.15ex\hbox{$/$}\kern-.53em\hbox{$k$}}
\newcommand{\slD}{\raise.15ex\hbox{$/$}\kern-.53em\hbox{$D$}}
\newcommand{\slC}{\raise.15ex\hbox{$/$}\kern-.53em\hbox{$C$}}
\newcommand{\slA}{\raise.15ex\hbox{$/$}\kern-.53em\hbox{$A$}}
\newcommand{\slSigma}{\raise.15ex\hbox{$/$}\kern-.53em\hbox{$\Sigma$}}
\newcommand{\slpartial}{\raise.15ex\hbox{$/$}\kern-.53em\hbox{$\partial$}}
\newcommand{\slcalP}{\raise.15ex\hbox{$/$}\kern-.63em\hbox{$\cal P$}}

\def\p{{\boldsymbol p}}
\def\q{{\boldsymbol q}}

\def\k{{\boldsymbol k}}

\def\x{{\boldsymbol x}}
\def\y{{\boldsymbol y}}
\def\X{{\boldsymbol X}}
\def\Y{{\boldsymbol Y}}

\def\r{{\boldsymbol r}}
\def\z{{\boldsymbol z}}
\def\v{{\boldsymbol v}}

\def\b{{\boldsymbol b}}

\catcode`\@=11

\newcount\@tempcntc
\def\@citex[#1]#2{\if@filesw\immediate\write\@auxout{\string\citation{#2}}\fi
  \@tempcnta\z@\@tempcntb\m@ne\def\@citea{}\@cite{%
        \@for\@citeb:=#2\do%
    {\@ifundefined{b@\@citeb}%
        {\@citeo\@tempcntb\m@ne\@citea%
                \def\@citea{,\penalty\@m\ }{\bf ?}\@warning%
                {Citation `\@citeb' on page \thepage \space undefined}}%
        {\setbox\z@\hbox{\global\@tempcntc0\csname b@\@citeb\endcsname\relax}
     \ifnum\@tempcntc=\z@ \@citeo\@tempcntb\m@ne%
       \@citea\def\@citea{,\penalty\@m}%
       \hbox{\csname b@\@citeb\endcsname}%
     \else%
      \advance\@tempcntb\@ne%
      \ifnum\@tempcntb=\@tempcntc%
      \else\advance\@tempcntb\m@ne\@citeo%
      \@tempcnta\@tempcntc\@tempcntb\@tempcntc\fi\fi}}\@citeo}{#1}}%

\def\@citeo{\ifnum\@tempcnta>\@tempcntb\else\@citea
  \def\@citea{,\penalty\@m}%
  \ifnum\@tempcnta=\@tempcntb\the\@tempcnta\else
   {\advance\@tempcnta\@ne\ifnum\@tempcnta=\@tempcntb \else
\def\@citea{--}\fi
    \advance\@tempcnta\m@ne\the\@tempcnta\@citea\the\@tempcntb}\fi\fi}

\catcode`\@=12

\begin{document}

\title{\bf High energy pA collisions\\
 in the color glass condensate approach\\
I.~Gluon~production and~the~Cronin~effect}
\author{Jean-Paul Blaizot$^{(1)}$, Fran\c cois Gelis$^{(1)}$, 
Raju Venugopalan$^{(2)}$}
\maketitle
\begin{center}
\begin{enumerate}
\item Service de Physique Th\'eorique\footnote{URA 2306 du CNRS.}\\
  B\^at. 774, CEA/DSM/Saclay\\
  91191, Gif-sur-Yvette Cedex, France
\item Physics Department\\
  Brookhaven National Laboratory\\
  Upton, NY 11973, USA
\end{enumerate}
\end{center}

\begin{abstract}
\noindent We study gluon production in high energy proton-nucleus
collisions in the semi-classical framework of the Color Glass
Condensate. We develop a general formalism to compute gluon fields in
covariant gauge to lowest order in the classical field of the proton
and to all orders in the classical field of the nucleus. The use of
the covariant gauge makes the diagrammatic interpretation of the
solution more transparnt.  $k_\perp$-factorization holds to this
order for gluon production -- Our results for the gluon distribution
are equivalent to the prior diagrammatic analysis of Kovchegov and
Mueller.  We also show that these results are equivalent to the
computation of gluon production by Dumitru and McLerran in the
Fock-Schwinger gauge.  We demonstrate how the Cronin effect arises in
this approach, and examine its behavior in the two extreme limits of
a)~no small-$x$ quantum evolution, and b)~fully saturated quantum
evolution. In both cases, the formalism reduces to Glauber's formalism
of multiple scatterings.  We comment on the possible implications of
this study for the interpretation of the recent results on
Deuteron-Gold collisions at the Relativistic Heavy Ion Collider
(RHIC).

\end{abstract}

\section{Introduction}
The physics of high energy proton/deuteron--nucleus collisions has
acquired new vigor with on-going experiments on Deuteron-Gold
collisions at center of mass energies per nucleon of $\sqrt{s} =
200$~GeV being conducted at Brookhaven's Relativistic Heavy Ion
Collider (RHIC). First results from these experiments have already
been submitted for publication and presented at
conferences~\cite{Debbe1,Frawl1,Stein1,Schwe1}.  In the near future,
CERN's Large Hadron Collider (LHC) will study proton-nucleus
collisions with center of mass energies per nucleon of $\sqrt{s}=5.5$
TeV~\cite{Accara1,Accara2,Bedjia1}.  When one considers this important
increase in the energy range for these collisions, it is interesting
to examine how results from lower energy proton-nucleus collisions are
modified both qualitatively and quantitatively as one goes to higher
energies.

At very high energies, or small $x$ values, the relevant parton
densities in the proton and in nuclei grow very rapidly. If these
densities are sufficiently large, the parton distributions may
saturate,~\cite{GriboLR1,MuellQ1,BlaizM1} leading to a qualitatively
different behavior of the distributions. Saturation will occur sooner
in nuclei than in protons because the large number of nucleons give
rise to an enhanced parton density in the transverse plane by a factor
$\sim A^{1/3}$.  Proton-nucleus collisions therefore, for a wide
energy range, provide an attractive physical environment wherein the
proton probe may be considered as a dilute parton gas with properties
that are believed to be well understood while the nucleus exists in a
novel, saturated high density state.  This latter state has been
called a Color Glass Condensate (CGC) and its distinctive features
have been extensively
explored~\cite{McLerV1,McLerV2,McLerV3,McLerV4,JalilKLW1,JalilKLW2,JalilKLW3,JalilKLW4,KovneM1,KovneMW3,Balit1,Kovch1,Kovch3,JalilKMW1,IancuLM1,IancuLM2,FerreILM1,IancuV1,IancuLM3,Muell4}.

In this work, we will formulate a description of high energy
proton-nucleus collisions within the CGC framework. In this framework,
the proton and the nucleus are effectively described as static random
sources of color charge on the light-cone. The leading contribution to
particle production is obtained by calculating the classical
Yang-Mills field created by these sources and then averaging over a
random distribution. Thus on is led to solve the Yang-Mills equations
in the presence of two number densities of color charges,
$\rho_p(\x_\perp)$ and $\rho_{_{A}}(\x_\perp)$ respectively for the
proton and the nucleus, localized on the light-cone~\footnote{The
  validity of this approach is not obvious a priori -- the motivations
  for this approach have been discussed extensively in the literature
  and we will not go into them here
  \cite{KovneMW1,KovneMW2,KovchR1,GyulaM1,KovchM3,KrasnV1,KrasnV2}. Here, it
  will be apparent only {\it a posteriori} -- from its success in
  reproducing known results and its predictive power.}. One has
\begin{eqnarray}
[D_\mu,F^{\mu\nu}]=J^\nu\, ,\nonumber
\end{eqnarray}
where 
\begin{eqnarray}
J^\nu_a=
g\delta^{\nu+}\delta(x^-)\rho_{p,a}(\x_\perp)
+
g\delta^{\nu-}\delta(x^+)\rho_{_{A},a}(\x_\perp)\; .
\label{eq:sources}
\end{eqnarray}
Operators calculated in this classical background field have to be
averaged over weight functionals $W_p$ and $W_{_{A}}$ -- representing
the distribution of the densities of color charges in the proton and
nucleus respectively.  One has then
\begin{eqnarray}
\langle O\rangle =\int [D\rho_p][D\rho_{_{A}}]
W_p[x_0^p,\rho_p]W_{_{A}}[{x_0^{_{A}}},\rho_{_{A}}] \,
O[\rho_p,\rho_{_{A}}] \, .
\label{eq:average}
\end{eqnarray}
This averaging procedure is essential in order to restore gauge
invariance since $O[\rho_p,\rho_{_{A}}]$ is computed in a particular
gauge. The arguments $x_0^p$ and $x_0^{_{A}}$ denote the scale in $x$
separating the large $x$ static sources from the small $x$ dynamical
fields. It is also through $W_p$ and $W_{_{A}}$ that quantum effects,
due to evolution with $x$ of the light-cone wave functions of the
target and projectile, are incorporated. The functionals $W_p$ and
$W_{_{A}}$ obey a Wilson renormalization group equation (often called
the JIMWLK
equation~\cite{JalilKLW1,JalilKLW2,JalilKLW3,JalilKLW4,KovneM1,KovneMW3,Balit1,Kovch1,Kovch3,JalilKMW1,IancuLM1,IancuLM2,FerreILM1}),
which governs their evolution with $x_0^p$ and $x_0^{_{A}}$. This
equation reduces to the well-known BFKL equation in the low density or
large transverse momentum limit.  In this paper, we will explicitly
show that the physical quantities relevant for proton-nucleus
collisions can be related to correlators of Wilson lines, the quantum
evolution of which can be studied thanks to the JIMWLK equation. 

The first part of this work, on gluon production in pA collisions,
will be presented in this paper.  The second part, on quark
production, will start from the solution of the Yang-Mills equations
obtained in the present paper and will be discussed in an accompanying
paper~\cite{BlaizGV2}.

This paper is organized as follows. In section
\ref{sec:YM-derivation}, we show how the Yang-Mills equations can be
truncated systematically in order to obtain a coupled set of equations
for the gauge fields to first order in the proton source density
$\rho_p$ and to all orders in the nuclear source density
$\rho_{_{A}}$. This approximation is valid as long as the parton
densities in the proton remains small: this is of course the case at
moderate collision energies, but also at very high collision energies
albeit in a more restricted kinematical domain (not too far away from
the fragmentation region of the proton). Gauge invariance is
preserved, by construction, in this truncation.  We explore the
structure of these truncated equations and introduce a convenient
diagrammatic expression for the gauge fields.  The equations are
solved explicitly in section \ref{sec:YM-solution}. We obtain a
compact expression for the gauge fields in terms of two distinct
Wilson lines whose coefficients are shown to be simply related to the
well known effective Lipatov
vertex~\cite{KuraeLF1,BalitL1,Lipat1,Duca1}.  Classical gluon
production is discussed in section \ref{sec:glue-prod}. It is shown
explicitly how the results can be written in a $k_\perp$-factorized
form as the product of two $k_\perp$-dependent distributions times a
factor proportional to the square of the Lipatov vertex. The results
obtained in this fashion are compared to the results of Dumitru and
McLerran \cite{DumitM1} in Fock-Schwinger gauge and are shown to be
identical.  In section \ref{sec:cronin}, we show how the Cronin effect
arises in simple models with and without quantum evolution of the
color charge densities. Readers interested in this topic alone may
skip directly to section \ref{sec:cronin}.  Interpreting our results
in light of the recent data from the BRAHMS experiment at RHIC for
both central and forward rapidities~\cite{Debbe1}, we find interesting
ramifications for our understanding of the physics from both d-Au and
Au-Au experiments at RHIC as well as for the future Heavy Ion
experiments at CERN's Large Hadron Collider (LHC).  Various technical
details are included in appendices A and B. Finally, we comment on the
relation of results in our approach to the Glauber multiple scattering
formalism.  Since this is somewhat outside the main thrust of this
paper, we discuss it in a self-contained fashion in appendix
\ref{sec:glauber}. We conclude in section \ref{sec:conclusions} and
set the stage for the results on quark production to be presented in
\cite{BlaizGV2}.

Before we proceed further, we should point out that several results on
gluon production in this paper are not new. They were first derived,
in covariant gauge, by Kovchegov and Mueller~\cite{KovchM3} and
subsequently refined in later papers~\cite{KovchT1,KharzKT1}. These
results were also derived, shortly after the work of Kovchegov and
Mueller, by Dumitru and McLerran in the Fock-Schwinger or radiation
gauge~\cite{DumitM1}. (We should also mention the work of Braun in a
somewhat different approach~\cite{Braun1}.) Our derivation, albeit
also in covariant gauge, follows a different tack from the Feynman
diagram technique of Kovchegov and Mueller. We solve the Yang-Mills
equations explicitly for the light-cone sources in
Eq.~\ref{eq:sources}. The expressions for the gauge fields are derived
explicitly and one observes how the Lipatov vertex arises in this
approach. We also show, for the first time, the exact equivalence
between the approach of Kovchegov and Mueller with that of Dumitru and
McLerran. Our results for the gauge fields are new and will be useful
in deriving the result for quark production, which will be discussed
in~\cite{BlaizGV2} (henceforth referred to as paper II).  Our
discussion of the Cronin effect is also, in part, known from previous
work by us and other
authors~\cite{DumitJ1,DumitJ2,GelisP1,GelisJ3,Kopel1,KharzLM1,KharzKT1,BaierKW1,JalilNV1,AlbacAKSW1}.
Our results in the ``super-saturated'' limit are new and help clarify
the interpretation of the RHIC data. The fact that one recovers the
Glauber multiple scattering picture in the case where the correlations
of the color sources are Gaussian is well
known~\cite{McLerV5,IvanoNSZZ1,AccarG1} -- however the formalism
discussed here provides a systematic way to go beyond the limitations
of the Glauber model by incorporating the non-Gaussian correlations
that may arise via quantum evolution.

\section{Yang-Mills equations}
\label{sec:YM-derivation}
The Color Glass Condensate is a classical effective field theory which
describes the physics of high energy, semi-hard processes in QCD. In
this approach, one first solves the classical Yang-Mills equations in
the presence of the light-cone sources $\rho_p$ and $\rho_{_{A}}$,
next one computes the observable of interest in this classical field,
and finally one averages over the sources $\rho_p$ and $\rho_{_{A}}$.
For proton-nucleus collisions, one requires the solution of the gauge
field to lowest order\footnote{Going beyond this approximation and
  solving the Yang-Mills equations to all orders in both sources has
  only been possible numerically so far
  \cite{KrasnV1,KrasnV2,KrasnNV1,KrasnNV2,Lappi1}.} in the source density
$\rho_p$ of the proton and to all orders in the nuclear source density
$\rho_{_{A}}$.  We shall write down in this section the truncated set
of Yang-Mills equations to this order. We next show that these
equations can be written in a compact form and explore their structure
and their diagrammatic content. Solutions to the Yang-Mills equations
will be discussed in the next section.

\subsection{Generalities and notations}
The classical Yang-Mills equations read\footnote{By
convention: $D^\mu\equiv\partial^\mu-ig A^\mu$.}:
\begin{equation}
[D_\mu,F^{\mu\nu}]=J^\nu\; .
\label{eq:eq-YM}
\end{equation}
$J^\nu$ is a color current which at lowest order in the sources
$\rho_p$ and $\rho_{_{A}}$ reads:
\begin{equation}
J^\nu_a=
g\delta^{\nu+}\delta(x^-)\rho_{p,a}(\x_\perp)
+
g\delta^{\nu-}\delta(x^+)\rho_{_{A},a}(\x_\perp)\; .
\label{eq:current-order-1}
\end{equation}
$\rho_p$ is the color source describing the proton, moving in the $+z$
direction at the speed of light. $\rho_{_{A}}$ is the color source
describing the nucleus, moving in the opposite light-cone direction.
These sources represent the number densities of color charges in the
proton and nucleus respectively, which are Lorentz contracted to delta
functions~\cite{KovneMW1,KovneMW2,KovchR1} -- $\delta(x^\pm)$.  In
general, the current $J^\nu_a$ receives higher order corrections in
$\rho_p, \rho_{_{A}}$ because it must be covariantly conserved:
\begin{equation}
[D_\nu,J^\nu]=0\; ,
\label{eq:eq-cons}
\end{equation}
and the produced gauge field has a feedback on the current itself.
Moreover, this system of equations is still under-determined because of
gauge invariance. Here, we choose the Lorenz gauge:
\begin{equation}
\partial_\mu A^\mu=0\; .
\end{equation}
Using this gauge condition, one can bring the Yang-Mills equations to
the following form
\begin{equation}
\square A^\nu=J^\nu+ig[A_\mu,F^{\mu\nu}+\partial^\mu A^\nu]\; .
\end{equation}
Since the commutator starts at least one order in $\rho_{p,_{A}}$ after
the gauge field itself, this form is appropriate for an expansion in
powers of the color sources.

In this section where we deal with truncations of the Yang-Mills
equations, it is convenient to introduce the following notations. For
any quantity $X$, let us denote by $X_{(m,n)}$ the term of order
$\rho_p^m\rho_{_{A}}^n$ in its expansion in powers of the
sources.  We therefore have:
\begin{equation}
X=\sum_{m=0}^{+\infty}\sum_{n=0}^{+\infty}X_{(m,n)}\; .
\end{equation}
Let us also denote by $X_{(0,\infty)}$ and $X_{(1,\infty)}$ the quantity $X$
at order 0 and 1 in $\rho_p$ and to all orders in $\rho_{_{A}}$, i.e.:
\begin{eqnarray}
&&X_{(0,\infty)}\equiv\sum_{n=0}^{+\infty} X_{(0,n)}\; ,\nonumber\\
&&X_{(1,\infty)}\equiv\sum_{n=0}^{+\infty} X_{(1,n)}\; .
\end{eqnarray}
Thus, for pA collisions in a kinematical regime where the proton
can be seen as a dilute object, we want to determine the component
$A_{(1,\infty)}^\nu$ of the gauge field.

\subsection{Order $\rho_p^0$}
Before we proceed to consider the field generated in proton-nucleus
collisions, let us first discuss the well-known case of the classical
field of a nucleus (before the collision).  If we first rewrite the
Yang-Mills and current conservation equations to order $\rho_p^0$, we
have:
\begin{eqnarray}
&&\square A_{(0,\infty)}^\nu=J^\nu_{(0,\infty)}
+ig[A_{(0,\infty)}{}_\mu,F_{(0,\infty)}^{\mu\nu}+\partial^\mu A_{(0,\infty)}^\nu]\; ,
\nonumber\\
&&\partial_\mu J_{(0,\infty)}^\mu=ig[A_{(0,\infty)}{}_\mu,J_{(0,\infty)}^\mu]\; .
\end{eqnarray}
This system of equations can be solved iteratively to the appropriate
order in $\rho_{_{A}}$. Indeed, at any given step, the commutators of
the r.h.s.  always involve the gauge field obtained at the previous
step. At order $\rho_{_{A}}^1$, we have simply:
\begin{eqnarray}
&&J^\mu_{(0,1)}=g\delta^{\mu-}\delta(x^+)\rho_{_{A}}(\x_\perp)\; ,\nonumber\\
&&\partial_\mu J^\mu_{(0,1)}=0\; ,\nonumber\\
&&\square A^\mu_{(0,1)}=J^\mu_{(0,1)}\; .
\end{eqnarray}
Note that the current conservation is automatically satisfied at this
order, because $J^\mu$ has only a $-$ component that does not depend
on $x^-$ (hence\footnote{$\partial^+\equiv \partial/\partial x^-$,
  $\partial^-\equiv \partial/\partial x^+$.} $\partial_\mu
J_{(0,1)}^\mu=\partial^+ J_{(0,1)}^-=0$). The solution of the
Yang-Mills equation is also trivial\footnote{This notation is a convenient abuse of language. One should of course read:
\begin{equation*}
\frac{1}{{\bs\nabla}_\perp^2}\rho_{_{A}}(\x_\perp)\equiv
\int d^2\y_\perp \Big<\x_\perp\Big|\frac{1}{{\bs\nabla}_\perp^2}\Big|\y_\perp\Big> \rho_{_{A}}(\y_\perp)\; .
\end{equation*}}:
\begin{equation}
A_{(0,1)}^\mu=
-g\delta^{\mu-}\delta(x^+)\frac{1}{{\bs\nabla}_\perp^2}\rho_{_{A}}(\x_\perp)\; .
\end{equation}
Note that the field strength corresponding to this solution has only
two non-vanishing components:
\begin{equation}
F_{(0,1)}^{i-}=-F_{(0,1)}^{-i}=\partial^i A^-_{(0,1)}\; .
\end{equation}

At the order $\rho_{_{A}}^2$, the equations are:
\begin{eqnarray}
&&\partial_\mu J_{(0,2)}^\mu=ig[A_{(0,1)}{}_\mu,J_{(0,1)}^\mu]\; ,\nonumber\\
&&\square A_{(0,2)}^\nu=J_{(0,2)}^\nu
+ig[A_{(0,1)}{}_\mu,F_{(0,1)}^{\mu\nu}+\partial^\mu A_{(0,1)}^\nu]\; .
\end{eqnarray}
Given the solution obtained at the first order, the commutator
$[A_{(0,1)}{}_\mu,J_{(0,1)}^\mu]$ that appears in the current
conservation equation vanishes, and the solution is $J_{(0,2)}^\mu=0$
if we impose that the current vanishes in the remote past. The same is
true of the commutator
$[A_{(0,1)}{}_\mu,F_{(0,1)}^{\mu\nu}+\partial^\mu A_{(0,1)}^\nu]$, so
that we have simply $\square A_{(0,2)}^\nu=0$. Again, if we impose
that the field vanishes in the remote past, we have $A_{(0,2)}^\nu=0$.
Therefore, the gauge field has no term of order
$\rho_p^0\rho_{_{A}}^2$. These arguments can be trivially extended to
any order in $\rho_{_{A}}$, so that the solution at order
$\rho_p^0\rho_{_{A}}^1$ is in fact valid to all orders in
$\rho_{_{A}}$:
\begin{eqnarray}
&&J^\mu_{(0,\infty)}
=J^\mu_{(0,1)}=\delta^{\mu-}\delta(x^+)\rho_{_{A}}(\x_\perp)\; ,
\nonumber\\
&&A^\mu_{(0,\infty)}=A_{(0,1)}^\mu=-g\delta^{\mu-}\delta(x^+)
\frac{1}{{\bs\nabla}_\perp^2}
\rho_{_{A}}(\x_\perp)\; .
\label{eq:rho0}
\end{eqnarray}
In section \ref{sec:YM-solution} and following, we will denote the
field of the nucleus alone simply $A_{_{A}}^\mu$ instead of
$A_{(0,1)}^\mu$. Similarly, the field of the proton alone will be
denoted $A_p^\mu$ instead of $A_{(1,0)}^\mu$.

\subsection{Order $\rho_p$}
We now come to the case of physical interest. 
At the order $\rho_p^1$, the equations we need to solve are the following:
\begin{eqnarray}
&&\square A_{(1,\infty)}^\nu=J^\nu_{(1,\infty)}\nonumber\\
&&\qquad\qquad\;
+ig[A_{(1,\infty)}{}_\mu,F_{(0,1)}^{\mu\nu}+\partial^\mu A_{(0,1)}^\nu]
+ig[A_{(0,1)}{}_\mu,F_{(1,\infty)}^{\mu\nu}+\partial^\mu A_{(1,\infty)}^\nu]
\; ,\nonumber\\
&&\partial_\mu J_{(1,\infty)}^\mu=
ig[A_{(1,\infty)}{}_\mu,J_{(0,1)}^\mu]
+ig[A_{(0,1)}{}_\mu,J_{(1,\infty)}^\mu]
\; .
\end{eqnarray}
Let us begin with the equation for the current. We can make the
reasonable assumption\footnote{The fact that we can find an exact
  solution of the Yang-Mills and current conservation equations under
  this assumption will justify it a posteriori.} that
$J_{(1,\infty)}^i=0$, i.e. the current $J^\mu_{(1,\infty)}$ is purely
longitudinal. This assumes that the sources that produce this current
do not undergo any recoil, i.e. that they are eikonal light-cone
sources. In addition, we require that the component
$J^+_{(1,\infty)}$, which is a correction from interactions with the
nucleus to the current produced by the proton, depends
locally\footnote{This means that $\rho_p(\x_\perp)$ cannot enter in
  $J^+_{(1,\infty)}(\x_\perp)$ via
  $(1/{\bs\nabla_\perp^2})\rho_p(\x_\perp)$.} on $\rho_p(\x_\perp)$
(as was $J^+_{(1,0)}$). Since in the r.h.s. of the current
conservation equation, the only term which is local in $\rho_p$ is the
term $ig[A_{(0,1)}{}_\mu,J_{(1,\infty)}^\mu]$, we can in fact split
this equation into separate equations for the $+$ and $-$ components
of the current, as follows:
\begin{eqnarray}
&&\partial^- J_{(1,\infty)}^+=ig[A_{(0,1)}^-,J_{(1,\infty)}^+]\; ,\nonumber\\
&&\partial^+ J_{(1,\infty)}^-=ig[A_{(1,\infty)}^+,J_{(0,1)}^-]\; .
\end{eqnarray}
We can then write more explicitly the evolution equation for each
component of the gauge field $A_{(1,\infty)}^\mu$, using all the
information we know from the field at the previous order. For
$A_{(1,\infty)}^+$, we get:
\begin{equation}
\square A_{(1,\infty)}^+=J_{(1,\infty)}^+ 
+ig[A_{(0,1)}^-,\partial^+ A_{(1,\infty)}^+]\; .
\end{equation}
For the transverse components $A_{(1,\infty)}^i$, we have the following
equation:
\begin{equation}
\square A_{(1,\infty)}^i=
ig[\partial^i A_{(0,1)}^-,A_{(1,\infty)}^+]
-ig[A_{(0,1)}^-,\partial^i A_{(1,\infty)}^+]
+2ig[A_{(0,1)}^-,\partial^+ A_{(1,\infty)}^i]\; .
\end{equation}
And finally, the equation for $A_{(1,\infty)}^-$ is:
\begin{eqnarray}
&&\square A_{(1,\infty)}^-=J_{(1,\infty)}^-
+ig[A_{(1,\infty)}^+,\partial^- A_{(0,1)}^-]
+2ig[A_{(1,\infty)}^i,\partial^i A_{(0,1)}^-]
\nonumber\\
&&\qquad\qquad\qquad\;\!
+ig[A_{(0,1)}^-,F_{(1,\infty)}^{+-}+\partial^+ A_{(1,\infty)}^-]\; .
\end{eqnarray}
This last equation can be made more explicit if we replace the
strength tensor by its expression (to order $\rho_p$ and all orders in
$\rho_{_{A}}$):
\begin{equation}
F_{(1,\infty)}^{+-}=
\partial^+ A_{(1,\infty)}^-
-\partial^- A_{(1,\infty)}^+
-ig[A_{(1,\infty)}^+,A_{(0,1)}^-]\; ,
\end{equation}
which leads to:
\begin{eqnarray}
&&\square A_{(1,\infty)}^-=J_{(1,\infty)}^-
+ig[A_{(1,\infty)}^+,\partial^- A_{(0,1)}^-]
-ig[A_{(0,1)}^-,\partial^- A_{(1,\infty)}^+]
\nonumber\\
&&\qquad\qquad\qquad\;\!
+(ig)^2[A_{(0,1)}^-,[A_{(1,\infty)}^+,A_{(0,1)}^-]]
\nonumber\\
&&\qquad\qquad\qquad\;\!
+2ig[A_{(1,\infty)}^i,\partial^i A_{(0,1)}^-]
+2ig[A_{(0,1)}^-,\partial^+ A_{(1,\infty)}^-]\; .
\end{eqnarray}

\subsection{Interpretation of the equations}
In order to make the connection between the solution of the Yang-Mills
equation and the perturbative diagrammatic expansion, it is useful to
interpret the equations derived in the previous section in terms of
Feynman diagrams. These equations can be simplified by first rewriting
all the commutators in terms of matrices of the adjoint
representation, as follows:
\begin{equation}
[A,B]_a=if^{abc}A_b B_c
=-\left(T^b\right)_{ca}A_b B_c
=\left[(A\cdot T) B\right]_a\; .
\end{equation}
We can then exploit this compact notation to rewrite the current
conservation and Yang-Mills equations of the previous subsection as
follows:
\begin{eqnarray}
&&(\partial^--ig A_{(0,1)}^-\cdot T)J_{(1,\infty)}^+=0\; ,\nonumber\\
&&\partial^+ J_{(1,\infty)}^-=ig(A_{(1,\infty)}^+\cdot T)J_{(0,1)}^-\; ,
\end{eqnarray}
and
\begin{eqnarray}
&&(\square-ig A_{(0,1)}^-\cdot T \partial^+)A_{(1,\infty)}^+
=
J_{(1,\infty)}^+\; ,
\nonumber\\
&&(\square-2ig A_{(0,1)}^-\cdot T \partial^+)A_{(1,\infty)}^i
=
-ig(A_{(0,1)}^-\cdot T)\partial^i A_{(1,\infty)}^+
\nonumber\\
&&\qquad\qquad\qquad\qquad\qquad\qquad\quad\;
+ig(\partial^i A_{(0,1)}^-\cdot T)A_{(1,\infty)}^+\; ,
\nonumber\\
&&(\square-2ig A_{(0,1)}^-\cdot T\partial^+)A_{(1,\infty)}^-
=
J_{(1,\infty)}^-\nonumber\\
&&\qquad\qquad\qquad\quad
-ig(A_{(0,1)}^-\cdot T)\partial^- A_{(1,\infty)}^+
+ig(\partial^- A_{(0,1)}^-\cdot T) A_{(1,\infty)}^+
\nonumber\\
&&\qquad\qquad\qquad\quad
-(ig)^2 (A_{(0,1)}^-\cdot T)^2 A_{(1,\infty)}^+
-2ig (\partial^i A_{(0,1)}^-\cdot T)A_{(1,\infty)}^i\; .
\label{eq:eqs-YM}
\end{eqnarray}
When writing these equations, we have collected in the l.h.s. all the
homogeneous terms, and kept the inhomogeneous terms in the r.h.s. One
then sees that the structure of these equations requires that we solve
them in a specific order: we need to first determine
$J_{(1,\infty)}^+$, then $A_{(1,\infty)}^+$, $A_{(1,\infty)}^i$,
$J_{(1,\infty)}^-$, and finally $A_{(1,\infty)}^-$. This is so because the
quantities previously determined enter as source terms in the
evolution equation for the latter quantities.

It is in fact easy to understand how the various components of the
gauge field are coupled in these Yang-Mills equations by studying the
structure of the gluon propagator in the color field $A_{(0,1)}^-$
created by the nucleus. Forgetting the source terms
$J_{(1,\infty)}^\pm$ which are irrelevant in this discussion, we can
formally write the field $A_{(1,\infty)}^\mu(x)$ at a ``time'' $x^+$
in terms of the field $A_{(1,\infty)}^\mu(y)$ at an earlier time
$x^+_0$ via the retarded gluon propagator:
\begin{eqnarray}
A_{(1,\infty)}^\mu(x)=\int\limits_{y^+=-\infty}\, dy^- d^2\y_\perp
\; G_{_{R}}{}^{\mu}{}_{\nu}(x,y)\,2\partial_y^+
A_{(1,\infty)}^\nu(y)\; .
\label{eq:ret-evol}
\end{eqnarray} 
where the integration is carried out on an hypersurface $y^+={\rm
  const}$ (taken here to be infinitely remote in the past).
$G_{_{R}}^{\mu\nu}(x,y)$ is the retarded gluon propagator (in the
presence of the background field $A_{(0,1)}^-$).
Eq.~(\ref{eq:ret-evol}) is not obvious and is derived in appendix
\ref{sec:green}. In the following discussion, it will be convenient to
consider the Fourier transform of the propagator, defined as follows:
\begin{equation}
G_{_{R}}^{\mu\nu}(q,p)\equiv\int d^4x\,d^4y\, 
e^{iq\cdot x}e^{-ip\cdot y} G_{_{R}}^{\mu\nu}(x,y)\; .
\end{equation}

\begin{figure}[htbp]
\begin{center}
\resizebox*{!}{2cm}{\includegraphics{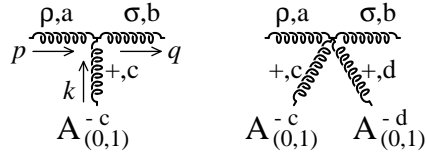}}
\end{center}
\caption{\label{fig:vertices} {\sl The  three gluon vertex 
    $\Gamma^{\sigma\rho+}_{bac}(q,p,k)$ and the  four gluon vertex
    $\Gamma^{\sigma\rho++}_{bacd}(q,p,k_1,k_2)$ that can be inserted
    on the gluon propagator.}}
\end{figure}

The full propagator can be obtained by multiple insertions of the
external field $A_{(0,1)}^-$ created by the nucleus. There are two
possible insertions, corresponding respectively to a three gluon
vertex or to a four gluon vertex, as illustrated in figure
\ref{fig:vertices}.
The free propagator, proportional to $g^{\mu\nu}$, does not mix the
various components $A_{(1,\infty)}^+$, $A_{(1,\infty)}^-$ and
$A_{(1,\infty)}^i$ of the gauge field. This property however does not
hold for the full propagator.  In order to see this, let us consider
the correction corresponding to the insertion of one external field
$A_{(0,1)}^-$, via the 3-gluon vertex:
\begin{eqnarray}
\delta G_{_{R}}^{\mu\nu}(q,p)=G{}_{_{R}}^0{}^{\mu}{}_{\sigma}(q)
\left(\Gamma^{\sigma\rho+}_{bac}(q,p,k)A_{(0,1)}^{-c}(k)\right)
G{}_{_{R}}^0{}_{\rho}{}^\nu(p)\; .
\end{eqnarray}
The vertex $\Gamma^{\sigma\rho+}_{bac}(q,p,k)$ is given by the usual
QCD Feynman rules in the covariant gauge, and reads (see figure
\ref{fig:vertices} for the direction of the momenta):
\begin{equation}
\Gamma^{\sigma\rho+}_{bac}(q,p,k)
=g f^{abc}\left[
g^{\rho\sigma}(p+q)^+
+
g^{\sigma+}(-q-k)^\rho
+
g^{\rho+}(k-p)^\sigma
\right]\; .
\end{equation}
More explicitly, its components are given by:
\begin{eqnarray}
&&\Gamma^{+++}_{bac}(q,p,k)=0\; ,\nonumber\\
&&\Gamma^{--+}_{bac}(q,p,k)=-g(p+q)^- f^{abc}\; ,\nonumber\\ 
&&\Gamma^{+-+}_{bac}(q,p,k)=\Gamma^{-++}_{bac}(q,p,k)=g p^+ f^{abc}\; ,\nonumber\\
&&\Gamma^{i++}_{bac}(q,p,k)=\Gamma^{+i+}_{bac}(q,p,k)=0\; ,\nonumber\\
&&\Gamma^{-i+}_{bac}(q,p,k)=-g (q+k)^i f^{abc}\; ,\nonumber\\
&&\Gamma^{i-+}_{bac}(q,p,k)=g (k-p)^i f^{abc}\; ,\nonumber\\
&&\Gamma^{ij+}_{bac}(q,p,k)=2g p^+ g^{ij} f^{abc}\; .
\end{eqnarray}
In these expressions, we have used the fact that $k^+=0$ (because the
background field $A_{(0,1)}^-$ does not depend on $x^-$).  This means
that all the components of $\delta G_{_{R}}^{\mu\nu}$ are non-zero,
except for $\delta G_{_{R}}^{++}$, $\delta G_{_{R}}^{+i}$ and $\delta
G_{_{R}}^{i+}$. Looking back at eq.~(\ref{eq:ret-evol}), this implies
that all the transitions $A_{(1,\infty)}^\nu\to A_{(1,\infty)}^\mu$
are allowed, except for the following transitions:
\begin{eqnarray}
&& A_{(1,\infty)}^-\to A_{(1,\infty)}^+\; ,\nonumber\\
&& A_{(1,\infty)}^-\to A_{(1,\infty)}^i\; ,\nonumber\\
&& A_{(1,\infty)}^i\to A_{(1,\infty)}^+\; ,
\end{eqnarray}
which cannot happen. The insertion of the 4-gluon vertex does not
change this conclusion. Indeed, the corresponding correction to the
propagator is given by:
\begin{eqnarray}
\delta G_{_{R}}^{\prime\ \mu\nu}(q,p)=G_0{}_{_{R}}^{\mu\sigma}(q)
\left(\Gamma^{\sigma\rho++}_{bacd}
A_{(0,1)}^-{}_c
A_{(0,1)}^-{}_d
\right)
G_0{}_{_{R}}^{\rho\nu}(p)\; ,
\end{eqnarray}
with a vertex that reads:
\begin{eqnarray}
\Gamma^{\sigma\rho++}
=ig^2 \delta^{\rho-}\delta^{\sigma-}(T^cT^d+T^d T^c)_{ab}\; .
\end{eqnarray}
Therefore, only the component $\delta G_{_{R}}^{\prime\ --}$ is non-zero.
This corresponds to a transition $A_{(1,\infty)}^+\to A_{(1,\infty)}^-$.

\noindent Our discussion thus far goes a long way towards explaining 
how the equations for $A_{(1,\infty)}^+$, $A_{(1,\infty)}^i$ and
$A_{(1,\infty)}^-$ are nested and what the couplings between the various
field components are. We note the following salient points:
\begin{itemize}
\item $A_{(1,\infty)}^+$ cannot be produced from $A_{(1,\infty)}^i$ or from
  $A_{(1,\infty)}^-$. Therefore, the equation for $A_{(1,\infty)}^+$ does
  not involve any other component of the field, and its only source
  term is the term in $J_{(1,\infty)}^+$. Moreover, the term in $ig
  (A_{(0,1)}^-\cdot T)\partial^+ A_{(1,\infty)}^+$ is the term that
  corresponds to the transition $A_{(1,\infty)}^+\to A_{(1,\infty)}^+$ by
  insertion of one power of the background field $A_{(0,1)}^-$, i.e.
  to the vertex $\Gamma^{+-+}$.
  
\item $A_{(1,\infty)}^i$ can be produced from $A_{(1,\infty)}^+$. Since
  there is no source term $J_{(1,\infty)}^i$, the inhomogeneous terms in
  the equation for $A_{(1,\infty)}^i$ corresponds to the vertex that is 
  responsible for the transition $A_{(1,\infty)}^+\to A_{(1,\infty)}^i$, i.e.
  $\Gamma^{i-+}$. The homogeneous term in this equation corresponds to
  the transition $A_{(1,\infty)}^i \to A_{(1,\infty)}^i$, i.e. to the vertex
  $\Gamma^{ii+}$.
  
\item In addition to the source term associated with the current
  $J_{(1,\infty)}^-$, $A_{(1,\infty)}^-$ can be produced from
  $A_{(1,\infty)}^+$ (vertices $\Gamma^{--+}$ and $\Gamma^{--++}$) or
  from $A_{(1,\infty)}^-$ (vertex $\Gamma^{-i+}$).  The homogeneous
  term in the equation for $A_{(1,\infty)}^-$ corresponds to the
  transition $A_{(1,\infty)}^-\to A_{(1,\infty)}^-$, i.e. to the
  vertex $\Gamma^{-++}$.
  
  One should note however that the equation for $A_{(1,\infty)}^-$, as
  written in eq.~(\ref{eq:eqs-YM}), does not quite fit with this
  interpretation. (Among other things, the homogeneous term is twice
  too large for this interpretation to hold.) However, this can be
  rescued by subtracting from the r.h.s. of this equation the quantity
  $ig[A_{(0,1)}^-,\partial_\mu A_{(1,\infty)}^\mu]$, which is zero
  according to the gauge condition we have chosen. After this
  subtraction has been performed, the equation for $A_{(1,\infty)}^-$
  becomes:
  \begin{eqnarray}
&&\!\!\!\!\!\!\!\!\!\!
(\square-ig A_{(0,1)}^-\cdot T\partial^+)A_{(1,\infty)}^-
=
J_{(1,\infty)}^-\nonumber\\
&&\!\!\!\!\!\!\!\!\!\!
-2ig(A_{(0,1)}^-\cdot T)\partial^- A_{(1,\infty)}^+
-ig(\partial^- A_{(0,1)}^-\!\cdot\! T) A_{(1,\infty)}^+
+g^2 (A_{(0,1)}^-\cdot T)^2 A_{(1,\infty)}^+
\nonumber\\
&&\!\!\!\!\!\!\!\!\!\!
-ig(A_{(0,1)}^-\cdot T)\partial^i A_{(1,\infty)}^i
-2ig (\partial^i A_{(0,1)}^-\cdot T)A_{(1,\infty)}^i\; .
  \end{eqnarray}
  One now  notes that all the terms in the equation correspond to
  the above interpretation.
\end{itemize}

\subsection{Diagrammatic content of the equations}
\begin{figure}[htbp]
\begin{center}
\resizebox*{!}{2cm}{\includegraphics{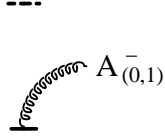}}
\end{center}
\caption{\label{fig:A0infty}{\sl 
  Diagrammatic representation of the field $A_{(0,1)}^-$. A solid
  boldface line represents one power of $\rho_{_{A}}$, and a dashed
  boldface line represents one power of $\rho_p$.}}
\end{figure}

At this stage, it is possible to sketch the Feynman diagrams that
correspond to the terms included into the solution of the Yang-Mills
equations at this order. The simplest case is the diagram
corresponding to the solution at order $\rho_p^0$. Since the field
$A_{(0,1)}^-$ is linear in the source $\rho_{_{A}}$, we can represent
it by the diagram of the figure \ref{fig:A0infty}.

\begin{figure}[h]
\begin{center}
\resizebox*{!}{2cm}{\includegraphics{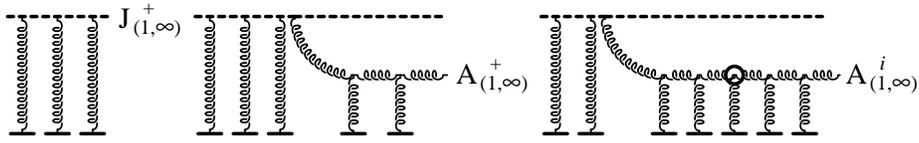}}
\end{center}
\caption{\label{fig:J1infty-plus}{\sl 
    Diagrammatic representation of the current $J_{(1,\infty)}^+$ and of
    the fields $A_{(1,\infty)}^+$ and $A_{(1,\infty)}^i$. The circled
    vertex is the vertex $\Gamma^{i-+}$ where the transition
    $A_{(1,\infty)}^+\to A_{(1,\infty)}^i$ takes place.}}
\end{figure}

The current $J_{(1,\infty)}^+$ is obtained by multiple scatterings of
the source $\rho_p$ in the field $A_{(0,1)}^-$, as illustrated by the
diagram on the left of figure \ref{fig:J1infty-plus}.

A typical contribution to $A_{(1,\infty)}^+$ is obtained by producing
a gluon from the current $J_{(1,\infty)}^+$, and then by letting this
gluon propagate through the background field $A_{(0,1)}^-$. This is
illustrated by the diagram in the middle of figure
\ref{fig:J1infty-plus}. The field $A_{(1,\infty)}^i$ has a source term
proportional to $A_{(1,\infty)}^+$ multiplied by a vertex
$\Gamma^{i-+}$ that enables the transition $A_{(1,\infty)}^+\to
A_{(1,\infty)}^i$. There cannot be more than one such vertex. Once
formed, the field $A_{(1,\infty)}^i$ propagates in the background
field $A_{(0,1)}^-$, as illustrated by the diagram on the right of
figure \ref{fig:J1infty-plus}.

\begin{figure}[htp]
\begin{center}
\resizebox*{!}{1.8cm}{\includegraphics{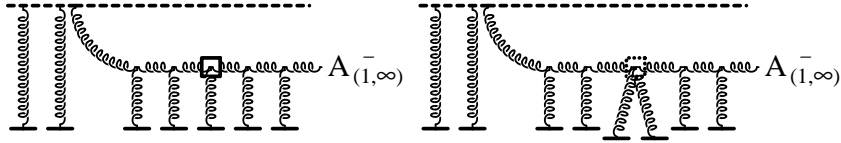}}
\end{center}
\caption{\label{fig:A1infty-minus-plus-1}{\sl 
    Contributions to $A_{(1,\infty)}^-$ by a direct transition
    $A_{(1,\infty)}^+\to A_{(1,\infty)}^-$. Left: via the 3-point vertex;
    the boxed vertex is the vertex $\Gamma^{--+}$. Right: via the
    4-point vertex; the dotted boxed vertex is the vertex
    $\Gamma^{--++}$.}}
\end{figure}

\begin{figure}[htp]
\begin{center}
\resizebox*{!}{1.8cm}{\includegraphics{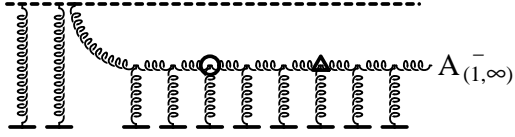}}
\end{center}
\caption{\label{fig:A1infty-minus-i}{\sl 
    Contributions to $A_{(1,\infty)}^-$ by a transition
    $A_{(1,\infty)}^+\to A_{(1,\infty)}^i\to A_{(1,\infty)}^-$. The triangular vertex is the vertex $\Gamma^{-i+}$.}}
\end{figure}

\begin{figure}[htp]
\begin{center}
\resizebox*{!}{1.8cm}{\includegraphics{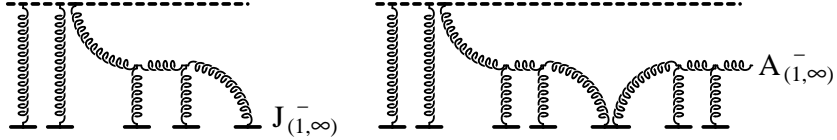}}
\end{center}
\caption{\label{fig:J1infty-minus}{\sl 
    Contribution to $J_{(1,\infty)}^-$, and the corresponding
    contribution to $A_{(1,\infty)}^-$.}}
\end{figure}

There are several sources for the field $A_{(1,\infty)}^-$. One
possible source is a direct transition $A_{(1,\infty)}^+\to
A_{(1,\infty)}^-$, as illustrated in figures
\ref{fig:A1infty-minus-plus-1} (there are a 3-gluon and a 4-gluon
vertex for this transition). Another contribution to
$A_{(1,\infty)}^-$ is due to the transition $A_{(1,\infty)}^+\to
A_{(1,\infty)}^i\to A_{(1,\infty)}^-$, as illustrated in figure
\ref{fig:A1infty-minus-i}. Finally, one last way to produce
$A_{(1,\infty)}^-$ is via the source $J_{(1,\infty)}^-$, which
requires the interaction of an $A_{(1,\infty)}^+$ with a source
$\rho_{_{A}}$. Note that this interaction is local in $\rho_{_{A}}$.
The diagrams contributing to $J_{(1,\infty)}^-$, as well as the
corresponding contribution to $A_{(1,\infty)}^-$, are illustrated in
figure \ref{fig:J1infty-minus}.

\section{Solution of the gauge field equations}
\label{sec:YM-solution}
Now that we have written down the truncated set of coupled equations
to lowest order $\rho_p$ in the source density of the proton and to
all orders in the source density of the nucleus, and understood the
structure of these equations both analytically and diagrammatically,
we are ready to solve the equations. This we will do in the order
prescribed in the previous section. Since at this point, there is no
risk of confusion, we can simplify our notations by removing the
indices that indicate the orders in $\rho_p$ and $\rho_{_{A}}$. From
now on, a quantity relative to the proton alone simply carries a
subscript $p$, a quantity relative to the nucleus alone has a
subscript $A$, and a quantity at order ($1,\infty$) has no subscript
at all. In particular, we have the following correspondence between
the notations of the previous section, and those of the rest of the
paper:
\begin{eqnarray}
&& A_{(0,1)}^\mu\longleftrightarrow A_{_{A}}^\mu\quad, \quad
A_{(1,0)}^\mu \longleftrightarrow A_{p}^\mu\; ,\nonumber\\
&& A_{(1,\infty)}^\mu\longleftrightarrow A^\mu\quad,\quad 
J_{(1,\infty)}^\mu\longleftrightarrow J^\mu\; .
\end{eqnarray}

\subsection{Expression for the current $J^+$}
In order to determine the source $J^+$, we have to solve the
equation:
\begin{equation}
(\partial^--ig A_{_{A}}^-\cdot T)J^+=0\; ,
\end{equation}
with the initial condition
$J^+(x^+=-\infty)=J_{p}^+=g\delta(x^-)\rho_p(\x_\perp)$.
The solution can be expressed as:
\begin{equation}
J^+(x^+,x^-,\x_\perp)=
g \,U(x^+,-\infty;\x_\perp)\,
\delta(x^-)\rho_p(\x_\perp)\; ,
\label{eq:J1pinfty}
\end{equation}
where
\begin{equation}
U(x_2^+,x_1^+;\x_\perp)\equiv {\cal P}_+ \exp\left[
ig \int_{x_1^+}^{x_2^+}dz^+
A_{_{A}}^-(z^+,\x_\perp)\cdot T
\right]
\end{equation}
is an adjoint Wilson line in the background field $A_{_{A}}^-$ (${\cal
  P}_+$ denotes the ordering along $x^+$). Here $A_{_{A}}^-$ can be
simply expressed in terms of the nuclear source density $\rho_{_{A}}$
through the expression given in eq.~(\ref{eq:rho0}).  Note that
eq.~(\ref{eq:J1pinfty}) depends locally on $\rho_p(\x_\perp)$, but
non-locally in $\rho_{_{A}}(\x_\perp)$ (because it depends on
$\rho_{_{A}}$ via the Weiz\"acker-Williams field $A_{_{A}}^-$).

\subsection{Expression for $A^+$}
The next thing to do is to determine the field $A^+$. It
obeys the following equation:
\begin{equation}
(\square-ig A_{_{A}}^-\cdot T \partial^+)A^+=J^+\; .
\end{equation}
We will go into the derivation of its solution in some detail because
several subtle issues arise (regarding smearing of sources and the
appearance of two different Wilson lines in the solution) that will be
relevant for the discussion in this paper and in \cite{BlaizGV2}.
Using eq.~(\ref{eq:A-solution}) in appendix \ref{sec:green}, we can
write the solution in terms of the source and of the initial condition
at $y^+=-\infty$ as follows:
\begin{eqnarray}
A^+(x)
&=&
\int d^4y\; G_{_{R}}(x,y)J^+(y)\nonumber\\
&+&\!\!\!\!\!\!
\int\limits_{y^+=-\infty} \!\!\!\!dy^-d^2\y_\perp\;
G_{_{R}}(x,y)2\partial_y^+ A^+(y)\; .
\label{eq:A+-1}
\end{eqnarray}
where $G_{_{R}}(x,y)$ is the retarded Green's function for the
operator $\square-ig A_{_{A}}^-\cdot T \partial^+$, defined by:
\begin{eqnarray}
&&(\square_x-ig A_{_{A}}^-(x)\cdot T \partial^+_x)G_{_{R}}(x,y)
=\delta^{(4)}(x-y)\; ,\nonumber\\
&&\lim_{x^+-y^+\to 0^+}G_{_{R}}(x,y)=\frac{1}{2}\theta(x^--y^-)\delta(\x_\perp-\y_\perp)\; ,
\nonumber\\
&&G_{_{R}}(x,y)\propto \theta(x^+-y^+)\; .
\label{eq:greenAplus}
\end{eqnarray}
The limit $x^+\to y^+$ can be obtained from the free retarded
propagator in the Feynman gauge\footnote{When the time difference goes
  to zero, the structure of the full propagator in $x^+$ becomes
  similar to that of the free propagator because there is not enough
  time for interactions with the background field to take place.}:
\begin{equation}
G_{_{R}}^0{}^{\mu\nu}(x,y)=
-g^{\mu\nu}
\int\frac{d^4k}{(2\pi)^4}
\frac{e^{-ik\cdot(x-y)}}{k^2+ik^+\epsilon}
=\frac{g^{\mu\nu}}{2\pi}\theta(x^+\!-\!y^+)\theta(x^-\!-\!y^-)\delta((x-y)^2)
\; .
\label{eq:free-ret-prop}
\end{equation}
When $x^+\to -\infty$, the gauge field must vanish because this is
before the projectiles start contributing to the field. Therefore, we
can take the second term in eq.~(\ref{eq:A+-1}) to be zero, and write
simply:
\begin{equation}
A^+(x)
=\int d^4y\; G_{_{R}}(x,y)J^+(y)\; .
\label{eq:A1infty-plus-1}
\end{equation}
The background field $A_{_{A}}^-$ is proportional to
$\delta(x^+)$, but some expressions are ambiguous unless we regularize
this delta function by giving it some small width. Therefore, we are
going to replace in intermediate steps the $\delta(x^+)$ by a regular
function $\delta_\epsilon(x^+)$ that is non-zero for
$x^+\in]0,\epsilon[$, and such that:
\begin{equation}
\int_{0}^{\epsilon}dx^+\,\delta_\epsilon(x^+)=1\; .
\end{equation}
Note that the necessity of such a regularization is a peculiarity of
the Lorenz gauge. Other oddities of this gauge will appear in the
course of our calculation, which will naturally disappear from all
physical quantities. Along with this regularization, we break down
eq.~(\ref{eq:A1infty-plus-1}) into three contributions, according to
the value of $y^+$:
\begin{eqnarray}
A^+(x)&=&
\int_{-\infty}^0 dy^+\int dy^-d\y_\perp\,G_{_{R}}(x,y)J^+(y)
\nonumber\\
&+&\int_0^\epsilon dy^+\int dy^-d\y_\perp\,G_{_{R}}(x,y)J^+(y)
\nonumber\\
&+&\int_\epsilon^{x^+} dy^+\int dy^-d\y_\perp\,G_{_{R}}(x,y)J^+(y)\; .
\label{eq:A1infty-plus-2}
\end{eqnarray}
We have assumed in this decomposition that $x^+ > \epsilon$. If this
were not the case, some of the terms may be dropped.

Let us first discuss the expressions for the current in
eq.~(\ref{eq:A1infty-plus-2}). The expressions for the propagator will
be discussed subsequently.  In the first term, ($y^+<0$), we can
replace the current $J^+$ by $J_{p}^+$. In the
intermediate term ($0<y^+<\epsilon$), we have\footnote{Since the
support of $A_{_{A}}^-$ is $0<x^+<\epsilon$, we have
$U(x^+,-\infty;\x_\perp)=U(x^+,0;\x_\perp)$.}:
\begin{equation}
J^+(y)=g\,U(y^+,0;\y_\perp)
\,\delta(y^-)\rho_p(\y_\perp)\; .
\end{equation}
It is clear that the $U$ which appears in this formula
corresponds to a color rotation of the incoming sources due to
interactions with the nucleus.  In the third term ($y^+>\epsilon$), we
have:
\begin{equation}
J^+(y)=
g\,U(\epsilon,0;\y_\perp)\,
\delta(y^-)\rho_p(\y_\perp)\; .
\end{equation}

We now consider the expressions for the propagators in the three terms
in eq.~(\ref{eq:A1infty-plus-2}). First we note the following
convolution property that can be obtained by applying twice
eq.~(\ref{eq:ret-evol}):
\begin{equation}
G_{_{R}}(x,y)=\int\limits_{z^+={\rm const}} dz^-d^2\z_\perp\,
G_{_{R}}(x,z)\,2\partial_z^+ G_{_{R}}(z,y)\; ,
\label{eq:prop-conv}
\end{equation}
valid for $y^+<z^+<x^+$ (see eq.~(\ref{eq:prop-conv})).  This may be
used in the first term of eq.~(\ref{eq:A1infty-plus-2}), where
$x^+>\epsilon$ and $y^+<0$, in order to split the propagator into
three pieces:
\begin{equation}
G_{_{R}}(x,y)=\int\limits_{v^+=\epsilon}\!\!dv^-d^2{\bs v}_\perp
\int\limits_{w^+=0}\!\!dw^-d^2{\bs w}_\perp\,
G_{_{R}}^0(x,v)2\partial_v^+ G_{_{R}}(v,w)
2\partial_w^+
G_{_{R}}^0(w,y)\; ,
\end{equation}
where we have used the fact that the propagator is a free propagator
when its two endpoints are both in $[-\infty,0]$ or in
$[\epsilon,+\infty]$. Similarly, for the second term of
eq.~(\ref{eq:A1infty-plus-2}) ($x^+>\epsilon$ and $0<y^+<\epsilon$),
we split the propagator into two pieces:
\begin{equation}
G_{_{R}}(x,y)=\int\limits_{v^+=\epsilon}dv^-d^2{\bs v}_\perp\,
G_{_{R}}^0(x,v)\,2\partial_v^+ G_{_{R}}(v,y)\; ,
\end{equation}
while in the third term ($x^+,y^+>\epsilon$) we simply have:
\begin{equation}
G_{_{R}}(x,y)=G_{_{R}}^0(x,y)\; .
\end{equation}
We see that the only non-trivial propagator we need is the propagator
connecting two points whose ``times'' $x^+$ and $y^+$ lie in the range
$[0,\epsilon]$. In this case, the retarded propagator simply reads:
\begin{equation}
G_{_{R}}(x,y)=\frac{1}{2}\theta(x^--y^-)\delta(\x_\perp-\y_\perp)
V(x^+,y^+;\y_\perp)
\; ,
\end{equation}
where:
\begin{equation}
V(x^+,y^+;\y_\perp)\equiv
{\cal P}_+\exp \left[i\frac{g}{2}\int\limits_{y^+}^{x^+}dz^+
A_{_{A}}^-(z^+,\y_\perp)\cdot T\right]
\; .
\end{equation}
This path ordered exponential is distinguished from the path ordered
exponential $U$ that we encountered previously by the factor of $1/2$
that appears in the exponent.  Notice further that
\begin{equation}
2\partial_x^+G_{_{R}}(x,y)=\delta(x^--y^-)\delta(\x_\perp-\y_\perp)
V(x^+,y^+;\y_\perp)
\; .
\label{eq:green1}
\end{equation}

With the expressions for the currents and the propagators in the three
distinct kinematical regions, we can now put together our expression for
the gauge field in eq.~(\ref{eq:A1infty-plus-2}). We first use the
expression in eq.~(\ref{eq:green1}) for the propagator and the simple
form for the current in the region $y^+<0$ to write the contribution
to $A^+$ from this region as
\begin{equation}
g\int\limits_{-\infty}^0dy^+ \int\limits_{y^-=0}d^2\y_\perp
\int\limits_{v^+=0}dv^-d^2\v_\perp
G_{_{R}}^0(x,v)V(\v_\perp)2\partial_v^+ G_{_{R}}^0(v,y)
\rho_p(\y_\perp)\; ,
\end{equation}
where we use a simpler notation for Wilson lines that run over the
whole longitudinal extent of the nucleus:
\begin{equation}
V(\x_\perp)\equiv 
V(\epsilon,0;\x_\perp)=
V(+\infty,-\infty;\x_\perp)\; .
\end{equation}
Wherever possible, we have taken the limit $\epsilon\to 0$. Note that
for the $1$, i.e. the identity term, in the $V$ path ordered
exponential, one can trivially convolute the two free propagators and
obtain:
\begin{equation}
g\int\limits_{-\infty}^0dy^+ \int\limits_{y^-=0}d^2\y_\perp
G_{_{R}}^0(x,y)\rho_p(\y_\perp)\; ,
\end{equation}
which is nothing else but a piece of the field $A_{p}^+$ of the
proton.

The contribution from the domain $y^+\in[0,\epsilon]$ goes to zero
when $\epsilon\to 0$.  The only other contribution to $A^+$ is the one
coming from the range $y^+>\epsilon$:
\begin{equation}
g\int\limits_{0}^{+\infty}dy^+ \int\limits_{y^-=0}d^2\y_\perp
G_{_{R}}^0(x,y)U(\y_\perp)\rho_p(\y_\perp)\; .
\end{equation}
The $1$ in the ordered exponential $U$ gives the second half of
$A_{p}^+$.

We can now combine the above results to  write the
complete solution for $A^+(x)$ in the region where $x^+>0$:
\begin{eqnarray}
A^+(x)&=&
g\!\!\int\limits_{-\infty}^0\!\!\!dy^+ \!
\!\!\!\!\int\limits_{y^-=0}\!\!\!\!d^2\y_\perp
\!\!\!\!\int\limits_{v^+=0}\!\!\!\!dv^-d^2\v_\perp
G_{_{R}}^0(x,v)V(\v_\perp)2\partial_v^+ G_{_{R}}^0(v,y)
\rho_p(\y_\perp)
\nonumber\\
&+&
g\!\!\int\limits_{0}^{+\infty}\!\!\!dy^+ \!\!\!\!
\int\limits_{y^-=0}\!\!\!\!d^2\y_\perp
G_{_{R}}^0(x,y)U(\y_\perp)\rho_p(\y_\perp)\; .
\label{eq:A+-3}
\end{eqnarray}
{\it Physically, this result made of only two terms suggests the
  following. In a collision at very high energy, the proton must emit
  the gluon either before or after hitting the nucleus. Emitting the
  gluon during the collision with the nucleus is very unlikely at high
  energy because of Lorentz contraction of longitudinal distances.
  (Such a process is suppressed by at least one power of the inverse
  collision energy.)}  Note however that, rather surprisingly, both
path ordered exponentials $U$ and $V$ contribute to the expression of
the gauge field. Naively, we would have believed only the contribution
from the color rotation of the sources, the $U$'s would have been
relevant; the $V$ contribution, from the particular form of the
propagator in this gauge, is unexpected.  We will see that our
intuitive expectations will be restored when we compute the gluon
distribution~\footnote{Albeit, further surprises await when we
  consider quark production~\cite{BlaizGV2}.}.

In this derivation of $A^+$, we rightly assumed throughout
that $x^+ >\epsilon$. However, for later use in the calculation of
$A^i$ and $A^-$, we need also the value of
$A^+(x)$ for $x^+\in[0,\epsilon]$. Only the term where the
gluon is emitted before the interaction of the proton with the nucleus
can contribute here, and the last free propagator does not appear. We
obtain,
\begin{eqnarray}
A^+(0<x^+<\epsilon)=
g\int\limits_{-\infty}^0 dy^+ 
\int\limits_{y^-=0}d^2\y_\perp
V(x^+,0;\x_\perp)G_{_{R}}^0(x,y)\rho_p(\y_\perp)\; .
\label{eq:A+-2}
\end{eqnarray}

It is also useful to compute the Fourier transform\footnote{We do not
  use a special symbol for Fourier transforms, since the context makes
  obvious what object we are dealing with.} of the field $A^+(x)$, for
which we obtain\footnote{When computing this Fourier transform, one
  must remember that eq.~(\ref{eq:A+-3}) is not valid for $x^+<0$. In
  the region $x^+<0$, we simply have $A^+(x)=A_{p}^+(x)$.}
\begin{eqnarray}
&&{ A}^+(q)
=
{ A}_{p}^+(q)
\nonumber\\
&&\!\!\quad+
\frac{ig}{(q^-+i\epsilon)(q^2+iq^+\epsilon)}
\int d^2\y_\perp e^{-i\q_\perp\cdot\y_\perp} 
\Big[U(\y_\perp)-1\Big]\rho_p(\y_\perp)\nonumber\\
&&\!\!\quad-\frac{ig\,2q^+}{q^2+iq^+\epsilon}
\!\int\! d^2\y_\perp d^2\v_\perp\frac{d^2\k_\perp}{(2\pi)^2}
e^{i(\k_\perp-\q_\perp)\cdot\v_\perp}
\frac{\big[V(\v_\perp)\!-\!1\big]}{\k_\perp^2}
e^{-i\k_\perp\cdot\y_\perp}\rho_p(\y_\perp)\, .\nonumber\\
&&
\end{eqnarray}
This expression can also be put in the more compact form
\begin{eqnarray}
{ A}^+(q)
&=&
{ A}_{p}^+(q)
\nonumber\\
&+&
\frac{ig}{q^2+iq^+\epsilon}\int\frac{d^2\k_{1\perp}}{(2\pi)^2}
\Bigg\{
\frac{-k_{1\perp}^2}{q^-+i\epsilon}
\big[U(\k_{2\perp})-(2\pi)^2\delta(\k_{2\perp})\big]
\nonumber\\
&&\qquad\qquad
+
2q^+
\big[V(\k_{2\perp})-(2\pi)^2\delta(\k_{2\perp})\big]
\Bigg\}
\frac{\rho_{1}(\k_{1\perp})}{k_{1\perp}^2}\; ,
\label{eq:A1compact}
\end{eqnarray}
with the following notations:
\begin{eqnarray}
&&\k_{2\perp}\equiv \q_\perp-\k_{1\perp}\; ,\nonumber\\
&&U(\k_\perp)
\equiv \int d^2\x_\perp e^{-i\k_\perp\cdot\x_\perp}
U(\x_\perp)\; ,\nonumber\\
&&V(\k_\perp)
\equiv \int d^2\x_\perp e^{-i\k_\perp\cdot\x_\perp}
V(\x_\perp)\;  ,\nonumber\\
&&\rho_p(\k_\perp)\equiv
\int d^2\x_\perp e^{-i\k_\perp\cdot\x_\perp} \rho_p(\x_\perp)\; .
\end{eqnarray}

When we expand the path ordered exponentials to first order in
$A_{_{A}}^-$, we recover the solution by Kovchegov and
Rischke~\cite{KovchR1} for the gauge field to lowest order in {\it
both} sources:
\begin{equation}
q^2{ A}_{(1,1)}^+(q)
=
-g^3\int\frac{d^2\k_{1\perp}}{(2\pi)^2}
\left[q^+-
\frac{k_{1\perp}^2}{q^-}
\right]
\frac{\rho_{_A}(\k_{2\perp})\cdot T}{k_{2\perp}^2}
\frac{\rho_p(\k_{1\perp})}{k_{1\perp}^2}
\; .
\end{equation}
Note that it only thanks to the fact that the Wilson line $V$ has a
factor $1/2$ in its exponent that this lowest order result is
correctly reproduced.

\subsection{Expression for $A^i$} 
Since $A^i(x^+=-\infty)=0$, the component
$A^i$ of the field can be expressed as:
\begin{equation}
A^i(x)=\int d^4v
D_{_{R}}(x,v)J^i(v)\; ,
\end{equation}
where $D_{_{R}}$ is the retarded Green's function of the
operator~\footnote{The factor of 2 in eq.~(\ref{eq:greenAi}) is what
  distinguishes $D_{_{R}}$ from the propagator $G_{_{R}}$ in the
  previous subsection -- see eq.~(\ref{eq:greenAplus}).}  $\square-2ig
A_{_{A}}^-\cdot T\partial^+$:
\begin{equation}
(\square_x-2ig A_{_{A}}^-(x)\cdot T\partial_x^+)
D_{_{R}}(x,y)=\delta^{(4)}(x-y)\; ,
\label{eq:greenAi}
\end{equation}
and where we denote
\begin{equation}
J^i(x)\equiv ig\Big[
(\partial^i A_{_{A}}^-(x)\cdot T)A^+(x)
-
(A_{_{A}}^-(x)\cdot T)\partial^i A^+(x)
\Big]\; .
\end{equation}
Note that $J^i(x)$ is non-zero only for $x^+\in[0,\epsilon]$ because
of the proportionality to the field $A_{_{A}}^-(x)$. The propagator
$D_{_{R}}$ is almost the same as $G_{_{R}}$: the only difference is
that the ordered exponential $V$ is replaced by a $U$
(with the same arguments) in the region of interaction with
$A_{_{A}}^-$. It is easy to check that the propagator $D_{_{R}}(x,v)$
connecting a point in the interaction region $(v^+\in[0,\epsilon])$
and a point after the interaction region $(x^+>\epsilon)$ is given by:
\begin{equation}
D_{_{R}}(x,v)=G_{_{R}}^0(x,v)U(\epsilon,v^+;\v_\perp)\; .
\end{equation}
We need also eq.~(\ref{eq:A+-2}), which we can rewrite as follows:
\begin{eqnarray}
A^+(x)&=&
g\,V(x^+,0;\x_\perp)
\int\limits_{-\infty}^0 dy^+ 
\int\limits_{y^-=0}d^2\y_\perp
G_{_{R}}^0(x,y)\rho_p(\y_\perp)
\nonumber\\
&=&
V(x^+,0;\x_\perp)
A_{p}^+(x)\; .
\end{eqnarray}

We can now put together all the ingredients in order to obtain,
\begin{eqnarray}
A^i(x)
&=&
ig\int\limits_0^\epsilon dv^+ dv^- d^2\v_\perp\;
G_0{}_{_{R}}(x,v)
U(\epsilon,v^+;\v_\perp)\nonumber\\
&&\qquad\quad\times
\big[
(\partial_v^i A_{_{A}}^-(v)\cdot T)V(v^+,0;\v_\perp)A_{p}^+(v)
\nonumber\\
&&\qquad\quad\;
-
(A_{_{A}}^-(v)\cdot T)\partial_v^i(V(v^+,0;\v_\perp)A_{p}^+(v))
\big]\; .
\end{eqnarray}
The Fourier transform of this expression is given by 
\begin{eqnarray}
{ A}^i(q)
&=&
\frac{ig}{q^2+iq^+\epsilon}\int\limits_0^\epsilon dv^+ dv^- d^2\v_\perp\;
e^{i q\cdot v}
U(\epsilon,v^+;\v_\perp)
\nonumber\\
&&\qquad\quad\times
\big[
(\partial_v^i A_{_{A}}^-(v)\cdot T)V(v^+,0;\v_\perp)A_{p}^+(v)
\nonumber\\
&&\qquad\quad\;
-
(A_{_{A}}^-(v)\cdot T)\partial_v^i(V(v^+,0;\v_\perp)A_{p}^+(v))
\big]\; .
\label{eq:Ai-1}
\end{eqnarray}
At this stage, in order to further simplify our result, we must find a
way to deal with the convolution of a $U$ and a
$V$. Using the results justified in appendix \ref{sec:B}, we
can write this expression as
\begin{eqnarray}
{ A}^i(q)
&=&
\!\!\frac{2}{q^2+iq^+\epsilon}
\int \!\!dv^-d^2\v_\perp\, e^{i(q^+v^--\q_\perp\cdot v_\perp)} 
\nonumber\\
&&\times\big\{
(V(\v_\perp)\!-\!U(\v_\perp))
\partial_v^i A_{p}^+(v)
+(\partial_v^i V(\v_\perp))A_{p}^+(v)\big\}\; .
\end{eqnarray}
A final simplification occurs after writing the terms on the r.h.s in
terms of their Fourier components,
\begin{equation}
{ A}^i(q)\!=\!\frac{2ig}{q^2+iq^+\epsilon}
\!\int\!\frac{d^2\k_{1\perp}}{(2\pi)^2}
\Big\{
k_1^i (V(\k_{2\perp})\!-\!U(\k_{2\perp}))
\!+\!
k_2^i V(\k_{2\perp})
\Big\}
\frac{\rho_p(\k_{1\perp})}{k_{1\perp}^2}\; .
\end{equation}
Expanding the time-ordered exponentials to first order in the
background field $A_{_{A}}^-$, we recover the lowest order solution of
Kovchegov and Rischke~\cite{KovchR1} for the transverse components of
the gauge field at that order,
\begin{equation}
q^2 { A}_{(1,1)}^i(q)
=-g^3\int\frac{d^2\k_{1\perp}}{(2\pi)^2}
\Big[k_2^i-k_1^i\Big]
\frac{\rho_{_A}(\k_{2\perp})\cdot T}{k_{2\perp}^2}
\frac{\rho_p(\k_{1\perp})}{k_{1\perp}^2}\; .
\end{equation}

\subsection{Expression for $A^-$}
One way of determining $A^-$ would be to solve its equation
of motion with the appropriate source terms. Since there are several
source terms, this procedure is somewhat lengthy. It is much easier
instead to obtain the final component of the field $A^\mu$
from the gauge condition. In Fourier space, the latter reads as
follows:
\begin{equation}
q^- { A}^+(q)
+
q^+ { A}^-(q)
-\q_\perp\cdot {{\bs A}}{}_\perp(q)=0\; .
\end{equation}
This further implies that 
\begin{eqnarray}
{ A}^-(q)
=\frac{1}{q^+}\left[
\q_\perp\cdot {{\bs A}}{}_\perp(q)
-q^- { A}^+(q)
\right]\; .
\end{eqnarray}

Substituting our results for $A^{+}$ and $A^i$
from the previous two subsections, we obtain,
\begin{eqnarray}
&&{ A}^-(q)=
\frac{ig}{q^2+iq^+\epsilon}
\int\frac{d^2\k_{1\perp}}{(2\pi)^2}
\Bigg\{
\left(\frac{k_{2\perp}^2-q_\perp^2}{q^+}\right)
\big[U(\k_{2\perp})-(2\pi)^2\delta(\k_{2\perp})\big]
\nonumber\\
&&\qquad\qquad+\left(2\frac{q_\perp^2}{q^+}-2q^-\right)
\big[V(\k_{2\perp})-(2\pi)^2\delta(\k_{2\perp})\big]
\Bigg\}
\frac{\rho_p(\k_{1\perp})}{k_{1\perp}^2}\; .
\end{eqnarray}

At order $0$ in $\rho_{_{A}}$, this expression is zero as expected. When we
expand the time-ordered exponentials to first order in this formula,
we again recover the known lowest order result~\cite{KovchR1},
\begin{equation}
q^2 { A}_{(1,1)}^-(q)=
-g^3\int\frac{d^2\k_{1\perp}}{(2\pi)^2}
\left[
\frac{k_{2\perp}^2}{q^+}-q^-
\right]
\frac{\rho_{_A}(\k_{2\perp})\cdot T}{k_{2\perp}^2}
\frac{\rho_p(\k_{1\perp})}{k_{1\perp}^2}\; .
\end{equation}

\subsection{Summary of results for $A^\mu$}
We can now collect our results for the components of the field
$A^\mu$ and examine these more closely.  By construction,
this solution obeys $\partial_\mu A^\mu=0$. We have also
checked that it has the correct lowest order limit. It is useful
to put this solution in the following compact form:
\begin{eqnarray}
{ A}^\mu(q)&=&{ A}_{p}^\mu(q)\nonumber\\
&+&\frac{ig}{q^2+iq^+\epsilon}
\int\frac{d^2\k_{1\perp}}{(2\pi)^2}
\Big\{
C_{_{U}}^\mu(q,\k_{1\perp})\, 
\big[U(\k_{2\perp})-(2\pi)^2\delta(\k_{2\perp})\big]
\nonumber\\
&&\qquad\qquad
+
C_{_{V}}^\mu(q)\, 
\big[V(\k_{2\perp})-(2\pi)^2\delta(\k_{2\perp})\big]
\Big\}
\frac{\rho_p(\k_{1\perp})}{k_{1\perp}^2}
\; ,
\label{eq:A1infty-final}
\end{eqnarray}
where we have defined the 4-vectors $C_{_{U,V}}^\mu$ as
follows, 
\begin{eqnarray}
&& 
C_{_{U}}^+(q,\k_{1\perp})\equiv -\frac{k_{1\perp}^2}{q^-+i\epsilon}
\;,\;
C_{_{U}}^-(q,\k_{1\perp})\equiv \frac{k_{2\perp}^2-q_\perp^2}{q^+}
\;,\; 
C_{_{U}}^i(q,\k_{1\perp})\equiv -2 k_1^i\; ,
\nonumber\\
&& 
C_{_{V}}^+(q)\equiv 2q^+ \quad,\quad 
C_{_{V}}^-(q)\equiv 2\frac{q_\perp^2}{q^+}-2q^-\quad,\quad 
C_{_{V}}^i\equiv 2 q^i\; .
\end{eqnarray}
These coefficients are simply related to the well known Lipatov
effective vertex~\cite{KuraeLF1,BalitL1,Lipat1,Duca1} $C_{_{L}}^\mu$
through the relation
$C_{_{L}}^\mu=C_{_{U}}^\mu+\frac{1}{2}C_{_{V}}^\mu$.  For any momentum
$q$, these 4-vectors satisfy
\begin{equation}
q\cdot C_{_{U}}=q\cdot C_{_{V}}=0\; ,
\end{equation}
which ensures that the covariant gauge condition is trivially
verified. Moreover, for a momentum $q$ {\sl on-shell}
($2q^+q^-=q_\perp^2$), we have the following additional properties:
\begin{eqnarray}
&&C_{_{U}}\cdot C_{_{V}}=C_{_{V}}^2=0\; ,\nonumber\\
&&C_{_{U}}^2=-4\frac{k_{1\perp}^2 k_{2\perp}^2}{q_\perp^2}=C_{_{L}}^2\; .
\end{eqnarray}
Remarkably, this ensures that when we use this solution in order to
compute the production of on-shell gluons, we get only the correlator
$\left<UU^\dagger\right>$ (the mixed correlators
$\left<UV^\dagger\right>$ and
$\left<VU^\dagger\right>$, as well as
$\left<VV^\dagger\right>$, do not appear). Moreover the
non-zero scalar product $C_{_{U}}^2$ is the square of the ordinary
Lipatov vertex.  This result suggests, as we will prove in the next
section, that we likely have the property of $k_\perp$-factorization
for gluon production in pA.

\section{Classical gluon production in pA collisions}
\label{sec:glue-prod}
\subsection{$k_\perp$-factorization in gluon production}
From the results derived in the previous section, it is fairly easy
now to derive the average multiplicity of gluons produced in pA
collisions, to all orders in the source describing the nucleus. Using
standard reduction formulas, the gluon production amplitude is given
in the classical approximation by:
\begin{equation}
{\cal M}_\lambda(\q)=
q^2 { A}^\mu(q)\epsilon_\mu^{(\lambda)}(\q)\; ,
\end{equation}
where $\epsilon_\mu^{(\lambda)}(\q)$ is the polarization vector for a
gluon of 3-momentum $\q$ in the polarization state $\lambda$. As
usual, what this formula tells us is that we need to Fourier transform
the gauge field, amputate the last free propagator (that's what the
$q^2$ prefactor does), put the result on shell, and project on the
appropriate polarization vector.  When we square this amplitude, we
can sum over all four polarization states including the two non
physical ones. Indeed, the property $q_\mu { A}^\mu=0$
ensures that the longitudinal polarization states are not going to
contribute to the result. We can therefore use the identity, 
\begin{equation}
\sum_{\lambda}\epsilon_\mu^{(\lambda)}(\q)\epsilon_\nu^{(\lambda)}{}^*(\q)=
-g_{\mu\nu}\; .
\end{equation}
The average number of gluons produced in a collision is given by the relation 
\begin{eqnarray}
\overline{N}_g&=&
\int \frac{d^3\q}{(2\pi)^3 2 E_q}
\sum_{\lambda}\left<\left|{\cal M}_\lambda(\q)\right|^2\right>
\nonumber\\
&=&
-g^2\int \!\frac{d^3\q}{(2\pi)^3 2 E_q}
\frac{d^2\k_{1\perp}}{(2\pi)^2}\frac{d^2\k_{1\perp}^\prime}{(2\pi)^2}
\;
\frac{C_{_{U}}(q,\k_{1\perp})\cdot C_{_{U}}(q,\k_{1\perp}^\prime)}
{k_{1\perp}^2k_{1\perp}^{\prime 2}}
\nonumber\\
&&\qquad\qquad\times
\left<
{\rho}_{p,a}^\dagger(\k_{1\perp})
{\rho}_{p,a^\prime}(\k_{1\perp}^\prime)
\right>
\left<
U^\dagger(\k_{2\perp})
U(\k_{2\perp}^\prime)
\right>_{aa^\prime}\; ,
\label{eq:glue-prod}
\end{eqnarray}
where $\left<\cdots\right>$ denotes the average over the color
sources.  Note that the field ${ A}_{p}^+(q)$ of the proton alone does
not contribute to on-shell gluon production. In order to rewrite this
expression in a more intuitive way, let us first express the
correlator of the proton sources $\rho_p$ in terms of the unintegrated
gluon distribution in the proton:
\begin{eqnarray}
&&g^2\left<{\rho}_{p,a}^\dagger(\k_{1\perp})
{\rho}_{p,a^\prime}(\k_{1\perp}^\prime)\right>=
\nonumber\\
&&\qquad=\frac{\delta^{aa^\prime}}{\pi d_{_{A}}}
\left[\frac{\k_{1\perp}+\k_{1\perp}^\prime}{2}\right]^2
\!\!\int\limits_{\X_\perp}\!\!
e^{i(\k_{1\perp}-\k_{1\perp}^\prime)\cdot \X_\perp}
\frac{d\varphi_p(\frac{\k_{1\perp}+\k_{1\perp}^\prime}{2}|\X_\perp)}
{d^2\X_\perp}\; .
\end{eqnarray}
The integration over $\X_\perp$ runs over the transverse profile of
the proton. $d_{_{A}}\equiv N^2-1$ is the dimension of the adjoint
representation of $SU(N)$, and $d\varphi_p/d^2\X_\perp$ is the number
of gluons per unit area and unit of transverse momentum in the proton
(i.e. the proton non-integrated gluon distribution per unit area).
Note that the impact parameter does not appear here simply because we
chose the origin of transverse coordinates at the center of the proton
($\b$ will therefore appear in the correlator involving the $U$'s).
This correlator can be approximated further if we realize that the
support for the integration over $\X_\perp$ is the transverse area of
the proton, a domain of typical size $1/\Lambda_{_{QCD}}$. This means
that the difference $\k_{1\perp}-\k_{1\perp}^\prime$, which is the
conjugate variable of $\X_\perp$, is at most of order
$\Lambda_{_{QCD}}$. Therefore, as long as we are interested in
transverse momenta much larger than $\Lambda_{_{QCD}}$, we can always
approximate $\k_{1\perp}\approx \k_{1\perp}^\prime$. In particular, we
can write:
\begin{equation}
g^2\left<{\rho}_{p,a}^\dagger(\k_{1\perp})
{\rho}_{p,a^\prime}(\k_{1\perp}^\prime)\right>=
\frac{\delta^{aa^\prime}}{\pi d_{_{A}}}\k_{1\perp}^2
\!\!\int\limits_{\X_\perp}\!\!
e^{i(\k_{1\perp}-\k_{1\perp}^\prime)\cdot \X_\perp}
\frac{d\varphi_p(\k_{1\perp}|\X_\perp)}{d^2\X_\perp}\; .
\label{eq:phi-proton}
\end{equation}
We will also neglect the difference $\k_{1\perp}-\k_{1\perp}^\prime$
in the other factors of eq.~(\ref{eq:glue-prod}).

Similarly, we can hide the correlator $\big< {
U}_{_{A}}^\dagger(\k_{2\perp}) U(\k_{2\perp}^\prime)
\big>_{aa^\prime}$ in a function $\varphi_{_{A}}$ that describes the
nucleus. By analogy with eq.~(\ref{eq:phi-proton}), we define it as
follows:
\begin{eqnarray}
&&\big<U^\dagger(\k_{2\perp})
U(\k_{2\perp}^\prime) \big>_{aa^\prime}=
\frac{g^2 N\delta_{aa^\prime}}{\pi d_{_{A}} }
\frac{1}{k_{2\perp}^2}
\nonumber\\
&&\qquad\qquad\times
\int d^2\Y_\perp e^{i(\k_{2\perp}^\prime-\k_{2\perp})\cdot \Y_\perp}\;
\frac{d\varphi_{_{A}}(\k_{2\perp}|\Y_\perp-\b)}{d^2\Y_\perp}\; ,
\label{eq:phi-nucleus}
\end{eqnarray}
where $\b$ is the impact parameter of the collision.  The
normalization of $\varphi_{_{A}}$ has been chosen so that it is
consistent with eq.~(\ref{eq:phi-proton}) in the case of a dilute
target. It is important to note that the function $\varphi_{_{A}}$ is
not the canonical unintegrated gluon distribution of the nucleus, i.e.
the expectation value of the number operator $a^\dagger_k a_k$: it
only coincides with the unintegrated gluon distribution at large
$\k_\perp$. The correct interpretation of the function
$\varphi_{_{A}}$ is that it is the square of the scattering amplitude
of a gluon on the target nucleus.

Neglecting the difference $\k_{1\perp}-\k_{1\perp}^\prime$ everywhere
except in the exponentials, and using the explicit expression of
$C_{_{U}}^2$, we obtain:
\begin{equation}
\overline{N}_g=\frac{16\alpha_s N}{\pi d_{_{A}} q_\perp^2}
\int \!\frac{d^3\q}{(2\pi)^3 2 E_q}
\frac{d^2\k_{\perp}}{(2\pi)^2}\;
\!\int\! d^2\X_\perp
\frac{d\varphi_p(\k_\perp|\X_\perp)}{d^2\X_\perp}
\frac{d\varphi_{_{A}}(\q_\perp\!-\!\k_\perp|\X_\perp\!-\!\b)}{d^2\X_\perp}\; .
\end{equation}
This is the $k_\perp$-factorized form of the gluon multiplicity for pA
collisions. We see that one can conveniently hide all the multiple
scattering effects in a generalization of the definition of the
unintegrated gluon distribution of the nucleus, the hard
(perturbative) part of the matrix element remaining the same as in pp
collisions.

This expression for proton-nucleus collisions was first derived by
Kovchegov and Mueller~\cite{KovchM3}, though not expressed in this
form (see also Kopeliovich et al.~\cite{Kopel1} and
Braun~\cite{Braun1}). It was re-written in this form first by
Kovchegov and Tuchin for deeply inelastic scattering~\cite{KovchT1}
and subsequently by Kharzeev, Kovchegov and Tuchin for the pA
case~\cite{KharzKT1}.

\subsection{Comparison with Dumitru and McLerran}
In this section, we compare our result for gluon production with the
result obtained by Dumitru and McLerran~\cite{DumitM1} in the
Fock-Schwinger gauge ($x^+ A^- + x^- A^+=0$) which is an interpolation
between two light-cone gauges.  It is simpler to do this comparison at
the level of the formula given by eq.~(\ref{eq:glue-prod}). The latter
may be rewritten as
\begin{eqnarray}
&&\frac{d\overline{N}_g}{d^2\q_\perp dy}
=-\frac{1}{16\pi^3}\int
\frac{d^2\k_{1\perp}}{(2\pi)^2}\frac{d^2\k_{1\perp}^\prime}{(2\pi)^2}
\;
\frac{C_{_{U}}(q,\k_{1\perp})\cdot C_{_{U}}(q,\k_{1\perp}^\prime)}
{k_{1\perp}^2k_{1\perp}^{\prime 2}}
\nonumber\\
&&\qquad\qquad\times
\left<
{\rho}_{p,a}^\dagger(\k_{1\perp})
{\rho}_{p,a^\prime}(\k_{1\perp}^\prime)
\right>
\left<
U^\dagger(\k_{2\perp})
U(\k_{2\perp}^\prime)
\right>_{aa^\prime}\; .
\end{eqnarray}
In the paper by Dumitru and McLerran, the gluon spectrum is given by
\begin{equation}
\left.\frac{d\overline{N}_g}{d^2\q_\perp dy}\right|_{_{D-ML}}
=\frac{2}{(2\pi)^2}{\rm tr}_{\rm c}\,\left<|a_1|^2+|a_2|^2\right>\; ,
\label{eq:N-dumitru}
\end{equation}
with coefficients $a_1$ and $a_2$ (converted to our notations)
defined to be
\begin{eqnarray}
&&a_1(\q_\perp)=\frac{e^{3i\pi/4}}{\sqrt{\pi}q_\perp}
\int d^2\x_\perp e^{-i\k_\perp\cdot\x_\perp}
\delta^{ij}{\alpha_1^{ai}(\x_\perp)}
\partial_x^j\Big(
t^{b} U(\x_\perp)_{ba}
\Big)\; ,\nonumber\\
&&a_2(\q_\perp)=\frac{e^{3i\pi/4}}{\sqrt{\pi}q_\perp}
\int d^2\x_\perp e^{-i\k_\perp\cdot\x_\perp}
\epsilon^{ij}\partial_x^j
\Big(
{\alpha_1^{ai}(\x_\perp)}
t^{b} U(\x_\perp)_{b a}
\Big)\; .
\end{eqnarray}
$t^b$ is a generator of the fundamental representation of $SU(N)$ and
${\rm tr}_{\rm c}$ denotes a trace of color matrices.  In these
equations, $\alpha_1^{ai}(\x_\perp)$ is defined by the relation
\begin{equation}
\alpha_1^{ai}(\x_\perp)\equiv 
-\frac{\partial_x^i}{{\bs\nabla}_\perp^2}\rho_{p,a}(\x_\perp)\; .
\end{equation}
It is convenient to replace $U(\x_\perp)$ and
$\alpha_1^{ai}(\x_\perp)$ by their Fourier transforms in the
expressions for $a_1$ and $a_2$:
\begin{eqnarray}
&&U(\x_\perp)=\int\frac{d^2\k_{2\perp}}{(2\pi)^2}
e^{i\k_{2\perp}\cdot\x_\perp} 
U(\k_{2\perp})\; ,\nonumber\\
&&\alpha_1^{ai}(\x_\perp)=\int\frac{d^2\k_{1\perp}}{(2\pi)^2}
e^{i\k_{1\perp}\cdot\x_\perp} 
\frac{ik_1^i}{k_{1\perp}^2}{\rho}_{p,a}(\k_{1\perp})\; .
\end{eqnarray}
We can thus write $a_1$ and $a_2$ as 
\begin{eqnarray}
&&a_1(\q_\perp)=\frac{e^{3i\pi/4}}{\sqrt{\pi}q_\perp}
\int\frac{d^2\k_{1\perp}}{(2\pi)^2}
\delta^{ij} \frac{k_1^i k_2^j}{k_{1\perp}^2}
{\rho}_{p,a}(\k_{1\perp}) t^{b} 
U(\k_{2\perp})_{ba}\; ,
\nonumber\\
&&a_2(\q_\perp)=\frac{e^{3i\pi/4}}{\sqrt{\pi}q_\perp}
\int\frac{d^2\k_{1\perp}}{(2\pi)^2}
\epsilon^{ij} \frac{k_1^i k_2^j}{k_{1\perp}^2}
\rho_{p,a}(\k_{1\perp}) t^{b} 
U(\k_{2\perp})_{ba}\; ,
\end{eqnarray}
where now we have implicitly $\k_{2\perp}=\q_\perp-\k_{1\perp}$.
Squaring these coefficients, taking the trace of their sum, averaging
over the sources and inserting the result in
eq.~(\ref{eq:N-dumitru}), we obtain:
\begin{eqnarray}
&&\left.\frac{d\overline{N}_g}{d^2\q_\perp dy}\right|_{_{D-ML}}
=
\frac{1}{4\pi^3 q_\perp^2}
\int\frac{d^2\k_{1\perp}}{(2\pi)^2}
\frac{d^2\k_{1\perp}^\prime}{(2\pi)^2}
\frac{\Big[
\delta^{ij}\delta^{kl}
+\epsilon^{ij}\epsilon^{kl}
\Big]
k_1^i k_2^j k_1^{\prime k} k_2^{\prime l}}{k_{1\perp}^2 k_{1\perp}^{\prime 2}}
\nonumber\\
&&\qquad\qquad\times
\left<
{\rho}_{p,a}^\dagger(\k_{1\perp})
{\rho}_{p,a^\prime}(\k_{1\perp}^\prime)
\right>
\left<
U^\dagger(\k_{2\perp})
U(\k_{2\perp}^\prime)
\right>_{aa^\prime}\; .
\end{eqnarray}
At this point, it is trivial (although a bit tedious) to verify the identity
\begin{equation}
-C_{_{U}}(q,\k_{1\perp})\cdot C_{_{U}}(q,\k_{1\perp}^\prime)
=
\frac{4}{q_\perp^2}
\Big[
\delta^{ij}\delta^{kl}
+\epsilon^{ij}\epsilon^{kl}
\Big]
k_1^i k_2^j k_1^{\prime k} k_2^{\prime l}
\; ,
\end{equation}
if the vector $q$ is on-shell and $k_1+k_2=k_1^\prime+k_2^\prime=q$.
This ends the proof that our result is equivalent to the result
obtained by Dumitru and McLerran. {\it Note that this equivalence
  works regardless of the model one chooses for the functional average
  over the hard sources.  One need not specify this average to prove
  our result, nor even assume translational invariance in the
  transverse plane.}

\section{Cronin effect in classical gluon production}
\label{sec:cronin}
The Cronin effect was discovered in proton-nucleus collisions in the
late 70's~\cite{AntreCFSK1,KlubePSAC1,CroniFSBM1}. The effect observed
was a hardening of the transverse momentum spectrum in proton-nucleus
collisions, relative to proton-proton collisions, that sets in at
transverse momenta of order $k_\perp \sim 1-2$~GeV, and disappears at
much larger $k_\perp$'s. A corresponding depletion was seen at low
transverse momenta, accompanied by a softening of the spectrum. At
that time, and indeed subsequently, the effect was interpreted as
arising from the multiple scatterings of partons from the proton off
partons from the nucleus \cite{KrzywEPS1}. As a result of such
scatterings, the partons acquire a transverse momentum kick, shifting
their momenta from lower to higher values, hereby causing the observed
respective depletion and enhancement. At high $k_\perp$, the higher
twist effects, which, in the language of perturbative QCD, are
responsible for multiple scattering~\cite{QiuS1,QiuS2} are suppressed
by powers of $k_\perp$.  The relative enhancement of the
cross-sections at moderate $k_\perp$'s should thus die away -- and
indeed, the data seemed to suggest as much.  Though a qualitative
understanding of the previously observed Cronin effect was suggested
by perturbative QCD, a quantitative agreement for all its features
(such as, for instance, the flavor dependence) is still lacking.

The high energy deuteron-nucleus collisions at RHIC (and at the LHC in
the near future) add another dimension to the Cronin effect, namely,
its energy dependence. At high collision energies, the projectile
probes the small $x$ partons in the nuclear wave-function, whose
distribution may be modified by the nuclear medium. As the energy
increases, there are effectively more of them to scatter from and this
should lead to an enhancement of the effect at higher
energies. Eventually, the number of gluons saturates, implying a
corresponding saturation of the effect of multiple
scatterings. However, what happens also at high energies is a change
in the spectrum and, as we shall see, that modifies qualitatively the
net effect of the multiple scatterings. These various types of
behaviors will be illustrated in this section.

First data from RHIC provide indications on how the Cronin effect is
modified with energy or, equivalently, with the rapidity. The $x$
values probed in these experiments, at $k_\perp \sim 2$~GeV, range
from $10^{-2}$ in the central rapidity region down to $10^{-4}$ at
very forward rapidities. The most dramatic result is that of the
BRAHMS experiment~\cite{Debbe1} which has taken data up to
pseudo-rapidities $\eta=3.2$~\footnote{The trends seen by BRAHMS are
also corroborated by PHOBOS, PHENIX and STAR in more limited kinematic
ranges~\cite{Frawl1,Stein1,Schwe1}. One can expect in the near future
results from more detailed studies of the forward region in
Deuteron-Gold collisions at RHIC from all experiments-in particular
from STAR and PHENIX.}. It is observed that the Cronin peak shrinks
rapidly with rapidity and at higher rapidity, one sees that there is a
significant suppression instead. Equally interesting is the centrality
dependence of the effect~\cite{Debbe1}.  At central rapidities, one
observes that the Cronin peak is enhanced in more central collisions,
while, for forward rapidities, the trend is reversed: more central
collisions at forward rapidities show a greater suppression than
peripheral collisions!
 
The Cronin effect has been studied in the Color Glass Condensate
framework by a number of authors. These studies range from
semi-quantitative~\cite{DumitJ1,DumitJ2,GelisJ3,KharzLM1,BaierKW1,JalilNV1}
to detailed numerical solutions of the Balitsky-Kovchegov
equation~\cite{AlbacAKSW1}.  There is a consensus among all these
authors that the Cronin effect is suppressed at forward rapidities.
It was shown early on by Dumitru, Jalilian-Marian, and one of
us~\cite{DumitJ1,DumitJ2,GelisJ3} that the McLerran-Venugopalan model
shows a greater Cronin enhancement for larger values of the saturation
scale $Q_s$. In particular, the location and the width of the Cronin
peak are roughly proportional to $Q_s$.  The saturation scale can be
larger for greater centralities or for smaller values of $x$. The
model therefore correctly predicts the greater enhancement for more
central collisions for $\eta=0$ at RHIC. However, it fails badly when
one goes to forward rapidities. Because $x$ is small, the saturation
scale $Q_s$ is larger and one gets more enhancement -- rather than the
observed suppression. One possible cause for the failure of the
McLerran-Venugopalan model is that it does not include the quantum
evolution of the parton distributions. In fact, it was noticed early
on that quantum corrections to the McLerran-Venugopalan model are
large~\cite{AyalaJMV1,AyalaJMV2}. The proper treatment of these
corrections was performed by several authors -- known collectively by
their acronyms as
JIMWLK~\cite{JalilKLW1,JalilKLW2,JalilKLW3,JalilKLW4,KovneM1,KovneMW3,Balit1,Kovch1,Kovch3,JalilKMW1,IancuLM1,IancuLM2,FerreILM1},
leading to the evolution equation for the weight functional
$W_{_{A}}[x_0^{_A},\rho_{_{A}}]$ for the sources $\rho_{_{A}}$. This
equation can equivalently be re-expressed as an infinite hierarchy of
ordinary differential equations for $n$-point correlators:
$\left<UU\right>\, \left<UUU\right>, \left<UUU\cdots\right>$, where
$U$ is a Wilson line\footnote{For more details,
see~\cite{IancuV1,IancuLM3,Muell4}.} in the background field created
by the source $\rho_{_{A}}$.  For the 2-point function
$\left<UU^\dagger\right>$, a closed equation exists in the large $N$
limit: it is the Balitsky-Kovchegov equation~\cite{Balit1,Kovch3}.
This equation reduces to the well-known BFKL equation in the low
density or large transverse momentum limit.  Both the BK equation and
the JIMWLK equation have been studied in analytical approximations
(see~\cite{LevinT1} and ~\cite{IancuLM2} respectively).  The BK
equation has been studied numerically~\cite{Braun2,ArmesB1} and has
been applied to a recent study of the Cronin effect~\cite{AlbacAKSW1}.
The JIMWLK equation has been studied analytically in the ``super
saturated'' small $x$ region~\cite{IancuIM2}. Detailed numerical
solutions have only become available recently~\cite{RummuW1}.
 
In this section, we will discuss the different features of the Cronin
effect in the Color Glass Condensate framework. In all cases studied,
the weight functional is a Gaussian. Then, the Cronin effect can be
equivalently understood in terms of independent multiple scatterings,
and the explicit connection between the present formalism and
Glauber's formalism of multiple scatterings is recalled in appendix
\ref{sec:glauber}.  We will first consider in detail the classical (no
quantum evolution) McLerran-Venugopalan model's predictions for the
Cronin effect. We next include quantum evolution ``naively'' by
introducing ``by hand'' an $x$ dependence in the saturation scale:
$Q_s\rightarrow Q_s(x)$, along the lines first suggested by
Golec-Biernat and W\"usthoff based on their parameterizations of the
HERA small-$x$ data~\cite{GolecW1,GolecW2}.  We show that this gives
the wrong behavior for the Cronin effect at forward rapidities. We
next consider analytical mean-field solutions of the JIMWLK equation
-- in a saturated regime where the system has no memory of its initial
conditions at small $x$. It was shown by Iancu, Itakura and
McLerran~\cite{IancuIM2}, that the weight functional of the sources in
this regime is also a Gaussian, albeit a non-local Gaussian
distribution. This Gaussian solution can be incorporated in our study
of the Cronin effect, and we find that if one is in this ``super
saturated'' regime already at $y=0$, the Cronin effect is suppressed.
Further evolution to $y=3$ increases the suppression, but not by a
great deal. We will discuss the implications of this study at the end
of this section.

\subsection{Plain MV model}
If we consider a large nucleus, there is an approximate translation
invariance of the correlator $\big<UU^\dagger\big>$, which implies
that the function $d\varphi_{_{A}}/d^2\X_\perp$ is approximately
independent of the location $\X_\perp$ in the transverse plane:
\begin{equation}
\frac{d\varphi_{_{A}}(\k_\perp|\X_\perp)}{d^2\X_\perp}
=
\frac{\pi^2 d_{_{A}}}{4 N g^2}
k_\perp^2 C(k_\perp)\; ,
\end{equation}
where we denote:
\begin{equation}
\int d^2\r_\perp e^{i\k_\perp\cdot\x_\perp}
\left<
U^\dagger(0)
U({\x_\perp})
\right>_{ab}
\equiv C(k_\perp)\delta_{ab}\; .
\end{equation}
This object is identical to the quantity $C(k_\perp)$ introduced in a
paper by one of us and Peshier~\cite{GelisP1,GelisP2}, except that it
is now a correlator of Wilson lines in the adjoint representation.
(This quantity, again for fundamental Wilson lines, was also studied
in~\cite{McLerV4,Venug1} where it was denoted $\gamma (k_\perp)$.)  In
the MV model, where the functional $W_{_A}$ is a Gaussian weight given
by
\begin{equation}
W_{_A}[\rho_{_{A}}]\equiv \exp\left[
 -\int d^2\x_\perp 
\frac{\rho_{_{A},a}(\x_\perp)\rho_{_{A},a}(\x_\perp)}{2\mu_{_{A}}^2}
\right]
\; ,
\label{eq:W-mv}
\end{equation}
this correlator can be computed in closed form. We get:
\begin{equation}
\left<
U^\dagger(0)
U({\x_\perp})
\right>_{ab}
=\delta_{ab}
\exp\left[-\frac{g^4 N \mu_{_{A}}^2}{2}\int d^2\y_\perp 
\big[
G_0(\y_\perp)-G_0(\y_\perp-\x_\perp)
\big]^2\right]\; ,
\end{equation}
where $G_0$ is the 2-dimensional massless propagator:
\begin{equation}
G_0(\x_\perp-\y_\perp)\equiv
\int\frac{d^2\k_\perp}{(2\pi)^2}
\frac{e^{i\k_\perp\cdot(\x_\perp-\y_\perp)}}{k_\perp^2}\; .
\end{equation}
In terms of the function $C$, the gluon spectrum reads:
\begin{equation}
\frac{d\overline{N}_g}{d^2\q_\perp dy}
=\frac{1}{16\pi^3 q_\perp^2}
\int \frac{d^2\k_\perp}{(2\pi)^2}
k_\perp^2 C(\k_\perp) \varphi_p(\q_\perp-\k_\perp)\; .
\label{eq:nbar-pA}
\end{equation}
At large transverse momentum $k_\perp$, we have the following behavior
for $C(k_\perp)$:
\begin{equation}
k_\perp^2 C(k_\perp)\approx
\frac{g^4 N \mu_{_{A}}^2}{k_\perp^2}\; ,
\end{equation}
which is the standard bremsstrahlung perturbative tail. Note that this
tail is proportional to $\mu_{_{A}}^2\propto A^{1/3}$, where $A$ is
the atomic number of the nucleus. However, at lower transverse
momentum ($k_\perp^2$ of the order of $Q_{s,_A}^2\sim g^4 N
\mu_{_{A}}^2$ or smaller), the $1/k_\perp^2$ growth stops and the
function $k_\perp^2 C(k_\perp)$ remains bounded when
$k_\perp\to 0$.
\begin{figure}[htbp]
\begin{center}
\resizebox*{!}{6cm}{\includegraphics{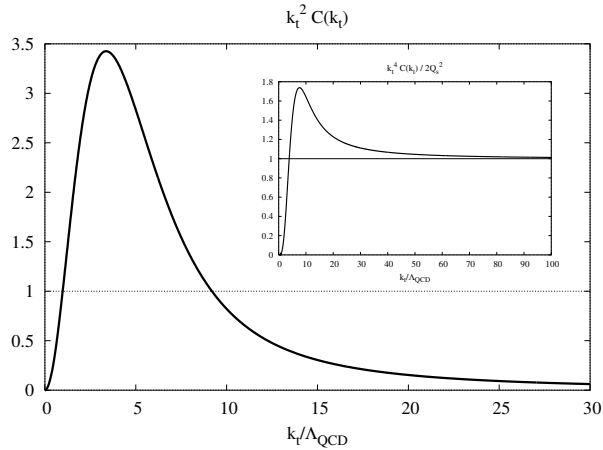}}
\end{center}
\caption{\label{fig:Ca} Behavior of 
  $\varphi(k_\perp)\propto k_\perp^2 C(k_\perp)$ in the
  McLerran-Venugopalan model. In the upper right corner is plotted the
  quantity $k_\perp^4 C(k_\perp)$, which shows that the tail of the
  function $\varphi(k_\perp)$ behaves like $1/k_\perp^2$. In this
  plot, we have taken $Q_{s,_A}^2\equiv g^4 N \mu_{_{A}}^2/2=25
  \Lambda_{_{QCD}}^2$, where $\Lambda_{_{QCD}}$ is the infrared cutoff
  (which sets the momentum scale).}
\end{figure}
This function was evaluated numerically in~\cite{GelisP1} and the
typical behavior of $k_\perp^2 C(k_\perp)$ is illustrated in
figure \ref{fig:Ca}.

\begin{figure}[htbp]
\begin{center}
\resizebox*{!}{6cm}{\includegraphics{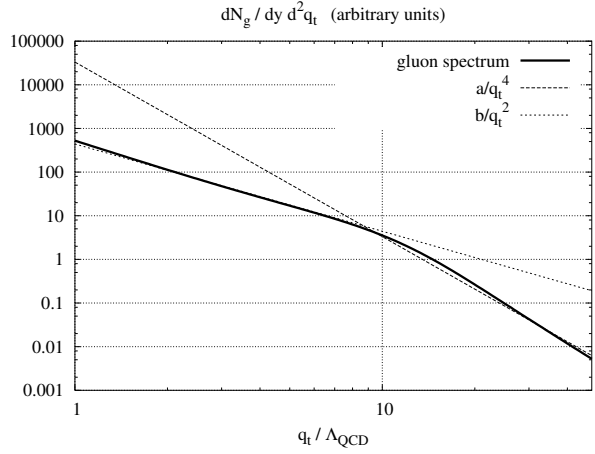}}
\end{center}
\caption{\label{fig:pA-spectrum} Numerical calculation of the spectrum
in pA collisions. We use $Q_{s,_{A}}/\Lambda_{_{QCD}}=5$ and
$Q_{s,p}/\Lambda_{_{QCD}}=2$. The dotted curves are fits to power
laws.}
\end{figure}

At low collision energy, we don't really know what the proton
unintegrated gluon distribution is because it is mostly a
non-perturbative quantity. In order to perform a quick numerical study
of eq.~(\ref{eq:nbar-pA}), we take the proton unintegrated gluon
distribution $\varphi_p(k_\perp)$ to be
proportional\footnote{Obviously, using a proton gluon distribution
  proportional to $k_\perp^2 C(k_\perp)$, i.e. to a correlator of
  Wilson lines, is a totally unjustified ansatz. Indeed, we have only
  proven $k_\perp$-factorization for a calculation performed at the
  lowest order in the source describing the proton. However, this
  ansatz for $\varphi_p$ has the expected large $k_\perp$ tail, and
  here one could simply see it as a way of introducing an infrared
  cutoff (necessary because the bremsstrahlung spectrum in
  $1/k_\perp^2$ is not integrable at low $k_\perp$). Moreover, what we
  take for $\varphi_p$ is not essential for the discussion of the
  Cronin effect, as it is mostly due to properties of the small-$x$
  wave function of the target.} to $k_\perp^2 C(k_\perp)$, with
the same function $C(k_\perp)$ but now evaluated with a much smaller
density $\mu_p^2$ (or saturation scale $Q_{s,p}^2$). We have now a
$C_{_{A}}(k_\perp)$ and a $C_p(k_\perp)$, corresponding respectively
to  $\mu_{_{A}}^2$ and  $\mu_p^2$, and the ratio of the two
saturation scales is taken to be:
\begin{equation}
\frac{\mu_{_{A}}^2}{\mu_p^2}=A^{1/3}\approx 6\; .
\end{equation}
Up to a constant normalization factor, we have
\begin{equation}
\left.\frac{d\overline{N}_g}{d^2\q_\perp dy}\right|_{pA}
\propto
\frac{1}{q_\perp^2}
\int \frac{d^2\k_\perp}{(2\pi)^2}
(\q_\perp-\k_\perp)^2 k_\perp^2 \;
C_{_{A}}(\k_\perp) C_p(\q_\perp-\k_\perp)\; ,
\label{eq:pA-spectrum}
\end{equation}
for proton-nucleus collisions, and
\begin{equation}
\left.\frac{d\overline{N}_g}{d^2\q_\perp dy}\right|_{pp}
\propto
\frac{1}{q_\perp^2}
\int \frac{d^2\k_\perp}{(2\pi)^2}
(\q_\perp-\k_\perp)^2 k_\perp^2 \;
C_p(\k_\perp) C_p(\q_\perp-\k_\perp)\; ,
\end{equation}
for pp collisions. The spectrum in pA collisions can be estimated
numerically using eq.~(\ref{eq:pA-spectrum}), and we have displayed
the result in figure \ref{fig:pA-spectrum}.
One can see clearly on this plot a change of behavior between a tail
in $1/q_\perp^4$ and an intermediate region where the slope of the
spectrum is $1/q_\perp^2$, as predicted by Dumitru and McLerran.

\begin{figure}[htbp]
\begin{center}
\resizebox*{!}{6cm}{\includegraphics{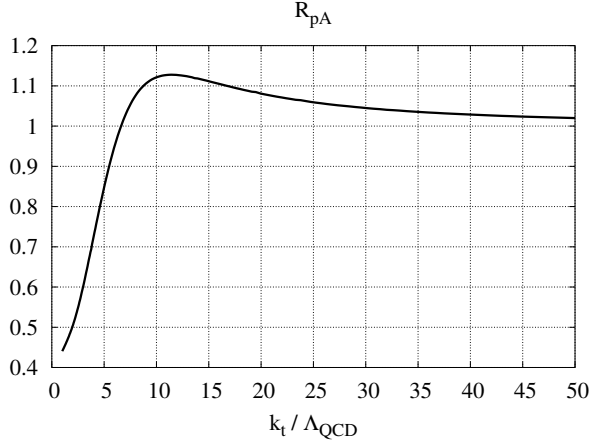}}
\end{center}
\caption{\label{fig:Rpa} The ratio $R_{pA}$ in the
McLerran-Venugopalan model, without quantum evolution. We use the
values $Q_{s,_{A}}/\Lambda_{_{QCD}}=4.9$ and $Q_{s,p}/\Lambda_{_{QCD}}=2$,
which corresponds to a ratio $\mu_{_{A}}^2/\mu_p^2\approx 6$.}
\end{figure}

We can also estimate the ``nuclear enhancement ratio'' $R_{pA}$ as
follows:
\begin{equation}
R_{pA}\equiv
\frac{\left.\frac{1}{N_{\rm coll}}
\frac{d\overline{N}_g}{d^2\q_\perp dy}\right|_{pA}}
{\left.\frac{d\overline{N}_g}{d^2\q_\perp dy}\right|_{pp}}
=
\frac{1}{A^{1/3}}
\frac{\int \frac{d^2\k_\perp}{(2\pi)^2}
(\q_\perp-\k_\perp)^2 k_\perp^2 \;C_{_{A}}(\q_\perp-\k_\perp) C_p(\k_\perp)}
{\int \frac{d^2\k_\perp}{(2\pi)^2}
(\q_\perp-\k_\perp)^2 k_\perp^2 \;C_p(\q_\perp-\k_\perp) C_p(\k_\perp)}\; .
\end{equation}
We have computed numerically the ratio of these integrals, and the
result is plotted in figure \ref{fig:Rpa}.
One can see clearly the Cronin effect as a small bump at
$q_\perp/\Lambda_{_{QCD}}\approx 11 $: the ratio is smaller than unity
at small momenta, and then remains larger than 1 at larger momenta.

When the saturation scale of the nucleus is much larger than that of
the proton, the location of the Cronin peak can be estimated in a
semi-quantitative way as follows. A good approximation of the behavior
of the function $C_{_{A}}(k_\perp)$ for transverse momenta $k_\perp$
comparable to the saturation scale or smaller is:
\begin{equation}
C_{_{A}}(k_\perp)\approx 
\frac{8\pi^2}{\overline{Q}_{s,_A}^2}
e^{-2\pi{k_\perp^2}/{\overline{Q}_{s,_A}^2}}\; ,
\label{eq:C-approx}
\end{equation}
with $\overline{Q}_{s,_A}^2\equiv Q_{s,_A}^2
\ln(Q_{s,_A}^2/\Lambda_{_{QCD}}^2)$ and $Q_{s,_A}^2\equiv 8\pi^2
\alpha_s^2 N \mu_{_{A}}^2$. Assuming that the ratio $R_{pA}$ can be
approximated by the ratio of the distributions in the
targets\footnote{This would be exact if the unintegrated gluon
  distribution of the projectile were proportional to a
  $\delta(\k_\perp)$. In practice, this is a reasonable approximation
  if the saturation scale in the nucleus target is much larger than
  that of the projectile.}
\begin{equation}
R_{pA}\approx \frac{C_{_{A}}(k_\perp)}{A^{1/3}C_p(k_\perp)}\; ,
\label{eq:R-approx}
\end{equation}
and that $C_p(k_\perp)\propto 1/k_\perp^4$ in the relevant
region\footnote{In the regime where the saturation scale in the proton
is much smaller than that in the nucleus, the bulk of the gluon
distribution of the nucleus corresponds to momenta that are in the
tail of the gluon distribution of the proton.}, we see that $R_{pA}$
has a maximum at $k_\perp=\overline{Q}_{s,_A}/\sqrt{\pi}$. The
apparent mismatch with the location of the maximum in figure
\ref{fig:Rpa} is due to an insufficient separation between $Q_{s,_A}$
and $Q_{s,p}$ (in addition to the fact that the two saturation scales
are not very large compared to $\Lambda_{_{QCD}}$).

\subsection{Quantum evolution introduced via $\mu_{_{A}}^2(x)$}
\begin{figure}[htbp]
\begin{center}
\resizebox*{!}{6cm}{\includegraphics{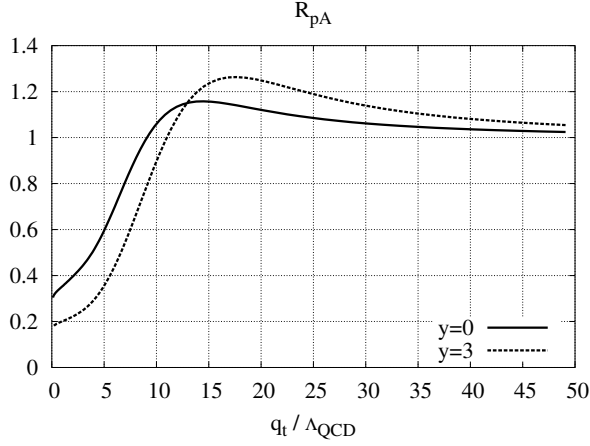}}
\end{center}
\caption{\label{fig:Rpa-evol1} The ratio $R_{pA}$ in the
McLerran-Venugopalan model, with quantum evolution introduced via the
$x$ dependence of the saturation scale.}
\end{figure}

The results of the previous subsection have been obtained in the plain
McLerran-Venugopalan model, which does not contain any quantum
evolution. This means that the quantities $\mu_{_{A}}^2, \mu_p^2$
(i.e. the saturation scales of the nucleus and of the proton) do not
depend on the momentum fraction $x$. One consequence is that the gluon
multiplicity is exactly boost invariant -- independent of $y$. A naive
extension of the previous calculations is to assume that the only
effect of quantum evolution is to change the values of $\mu_{_{A}}^2$
and $\mu_p^2$ by making them $x$ dependent, while preserving the local
Gaussian nature of the functional weight. Following a popular
parameterization, originally due to Golec-Biernat and
W\"usthoff~\cite{GolecW1,GolecW2}, we parametrize the saturation scale
as
\begin{eqnarray}
&&Q_{s,p}^2/Q_0^2 =\left(\frac{x_0}{x}\right)^\lambda\; ,\nonumber\\
&&Q_{s,_{A}}^2/Q_0^2 =A^{1/3}\left(\frac{x_0}{x}\right)^\lambda\; ,
\end{eqnarray}
with $Q_0\equiv 1$~GeV.  We use the values of $x_0$ and $\lambda$
obtained in the recent paper by Iancu, Itakura and Munier:
$x_0=0.67\cdot 10^{-4}$ and $\lambda=0.25$ from their fits to the HERA
$F_2$ data~\cite{IancuIM3}.  The values of $x$ in the proton and in
the nucleus are given by:
\begin{eqnarray}
&&x_p=\frac{q_\perp}{\sqrt{s}}e^{+y}\; ,\nonumber\\
&&x_{_{A}}=\frac{q_\perp}{\sqrt{s}}e^{-y}\; ,
\end{eqnarray}
where $\sqrt{s}=200\,{\rm GeV}$ is the energy per nucleon in the
collision. For definiteness, we set the infrared cutoff to
$\Lambda_{_{QCD}}=200\,{\rm MeV}$. We then simply repeat the previous
calculation with the following formula for $R_{pA}$:
\begin{equation}
R_{pA}
=
\frac{1}{A^{1/3}}
\frac{\int \frac{d^2\k_\perp}{(2\pi)^2}
(\q_\perp-\k_\perp)^2 k_\perp^2 \;
C_{_{A}}(x_{_{A}},\k_\perp) C_p(x_p,\q_\perp-\k_\perp)}
{\int \frac{d^2\k_\perp}{(2\pi)^2}
(\q_\perp-\k_\perp)^2 k_\perp^2 \;
C_p(x_{_{A}},\k_\perp) C_p(x_p,\q_\perp-\k_\perp)}\; .
\end{equation}
The results of the numerical evaluation of this ratio are displayed in
figure \ref{fig:Rpa-evol1}.

The characteristic Cronin bump is still present both at $y=0$ and at
$y=3$.  The primary difference is that at larger rapidity the maximum
is more pronounced and is pushed towards larger momenta. This is a
consequence of the fact that it is the $Q_s$ of the nucleus that
determines the location of the maximum to a large extent (see
eqs.~(\ref{eq:C-approx}) and (\ref{eq:R-approx})).  Similarly, more
central collisions would give a greater enhancement as well, since they
correspond to a larger $Q_s$.  This picture is consistent with the
centrality dependence observed at $y=0$ in Deuteron-Gold collisions at
RHIC.  It is however in complete disagreement, at forward rapidities,
with the observation made recently by the BRAHMS
experiment~\cite{Debbe1}.

\subsection{Non-local Gaussian distribution}
\label{sec:non-local}
One can generalize the previous model for the correlator
$\big<U^\dagger(0)U(\x_\perp)\big>_{ab}$ by allowing
transverse non-localities in the Gaussian functional $W_{_A}$. This
amounts to generalizing eq.~(\ref{eq:W-mv}) by allowing $\mu_{_A}^2$ to
depend not only on $x$ but also on the transverse coordinates:
\begin{equation}
W_{_A}[x,\rho_{_A}]\equiv \exp\left[
 -\int d^2\x_\perp d^2\y_\perp
\frac{\rho_{_{A},a}(\x_\perp)\rho_{_{A},a}(\y_\perp)}{2\mu_{_A}^2(x,\x_\perp-\y_\perp)}
\right]\; ,
\label{eq:W-non-local}
\end{equation}
so that the average of a product of two $\rho$ is given by
\begin{equation}
\left<
\rho_{_{A},a}(\x_\perp)\rho_{_{A},b}(\y_\perp)\right>=\delta_{ab}\mu_{_A}^2(x,\x_\perp-\y_\perp)\; .
\end{equation}
In this model, we have assumed that the nucleus is invariant by
translation in the transverse direction, which implies that
$\mu_{_A}^2$ depends only on the difference $\x_\perp-\y_\perp$. For
instance, this extension can be used to incorporate effects of color
charge neutralization \cite{LamM1} by making the function
$\mu_{_A}^2(x,\x_\perp-\y_\perp)$ vanish at distances larger than a
certain scale. In this model, one can still calculate in closed form
the correlator of two Wilson lines:
\begin{eqnarray}
&&\left<U^\dagger(0)U(\x_\perp)\right>_{ab}
=
\delta_{ab}
\exp\Big[
-\frac{g^4N}{2}\int d^2\y_\perp d^2\z_\perp\,\mu_{_A}^2(x,\y_\perp-\z_\perp)
\nonumber\\
&&\qquad\qquad\qquad\qquad\times
\big(G_0(\y_\perp)-G_0(\y_\perp-\x_\perp)\big)
\big(G_0(\z_\perp)-G_0(\z_\perp-\x_\perp)\big)
\Big]
\, ,\nonumber\\
&&
\end{eqnarray}
or by going to Fourier space,
\begin{eqnarray}
\left<U^\dagger(0)U(\x_\perp)\right>_{ab}
=\delta_{ab}
\exp\Big[
-\frac{g^4N}{2\pi}
\int\limits_0^{+\infty}\frac{dk_\perp}{k_\perp^3}(1-J_0(k_\perp x_\perp))
\mu_{_A}^2(x,k_\perp)
\Big]\; ,
\nonumber\\
\label{eq:UUdagger}
\end{eqnarray}
with
\begin{equation}
\mu_{_A}^2(x,k_\perp)\equiv\int d^2\x_\perp e^{i\k_\perp\cdot\x_\perp}
\mu_{_A}^2(x,\x_\perp)\; .
\end{equation}
(We have assumed rotational invariance in the transverse plane. This
Fourier transform therefore depends only on
$k_\perp=|\k_\perp|$).

Such a non-local Gaussian distribution of color sources has in fact
been predicted by Iancu, Itakura, McLerran~\cite{IancuLM2}, as a
mean-field asymptotic solution of the JIMWLK evolution equation for
the deeply saturated regime. In this asymptotic regime, the color
neutralization is controlled by saturation physics and therefore
happens at the scale $Q_{s}$ instead of $\Lambda_{_{QCD}}$, and the
function $\mu^2$ is given by:
\begin{equation}
\frac{g^4 N}{2\pi}\mu_{_A}^2(x,k_\perp)\equiv \frac{2}{\gamma c}
k_\perp^2 \,\ln\left(1+\Bigg(\frac{Q_{s,_A}^2(x)}{k_\perp^2}\Bigg)^\gamma\right)\; .
\label{eq:edmond-mu2}
\end{equation}
In this equation, $\gamma$ is some anomalous dimension ($\gamma\approx
0.64$ for BFKL) and $c\approx 4.84$. Before going on, it is
instructive to discuss the unintegrated gluon distribution
$\varphi(x,k_\perp)\propto k_\perp^2 C(x,k_\perp)$ for this model. The
function $C(x,k_\perp)$ can be expanded in powers of
$\mu_{_A}^2(x,k_\perp)$, which corresponds to an expansion in the
number of collisions (see the appendix \ref{sec:glauber}):
\begin{equation}
C(x,k_\perp)= C_1(x,k_\perp)+C_2(x,k_\perp)+\cdots\; ,
\end{equation}
with\footnote{Note that $\varphi(x,k_\perp)$ depends on $x$ and $k_\perp$
only via the ratio $k_\perp/Q_s(x)$.}
\begin{equation}
C_1(x,k_\perp)=g^4 N \frac{\mu_{_A}^2(x,k_\perp)}{k_\perp^4}
=\frac{4\pi}{\gamma c}
\frac{1}{k_\perp^2}
\ln\left(1+\Bigg(\frac{Q_{s,_A}^2(x)}{k_\perp^2}\Bigg)^\gamma\right)\; ,
\label{eq:C1}
\end{equation}
and
\begin{equation}
C_2(x,k_\perp)=
\frac{g^8 N^2}{2}\!\!\!
\int\!\frac{d^2\p_\perp}{(2\pi)^2}
\!\left[\!
\frac{\mu_{_A}^2(x,p_\perp)}{p_\perp^4}
\frac{\mu_{_A}^2(x,\left|\k_\perp\!-\!\p_\perp\right|)}{(\k_\perp-\p_\perp)^4}
\!-\!2
\frac{\mu_{_A}^2(x,k_\perp)}{k_\perp^4}
\frac{\mu_{_A}^2(x,p_\perp)}{p_\perp^4}\!
\right]\!\, .
\label{eq:C2}
\end{equation}
One can note that with the $\mu_{_A}^2(x,p_\perp)$ given in
eq.~(\ref{eq:edmond-mu2}), this quantity is infrared finite.  A
numerical study of this 2-scattering correction shows that it is
negative at low $k_\perp$ and positive at large $k_\perp$. Therefore,
the property that rescatterings move particles from low $k_\perp$ to
high $k_\perp$ regions of the spectrum holds also in this model.
However, since in this model the spectrum has a smaller slope at large
$k_\perp$ ($k_\perp^{-2(1+\gamma)}$ with $\gamma\approx 0.64$) than
the MV model spectrum ($k_\perp^{-4}$), this effect of rescatterings
is much less important.
\begin{figure}[htbp]
  \begin{center} \resizebox*{!}{6cm}{\includegraphics{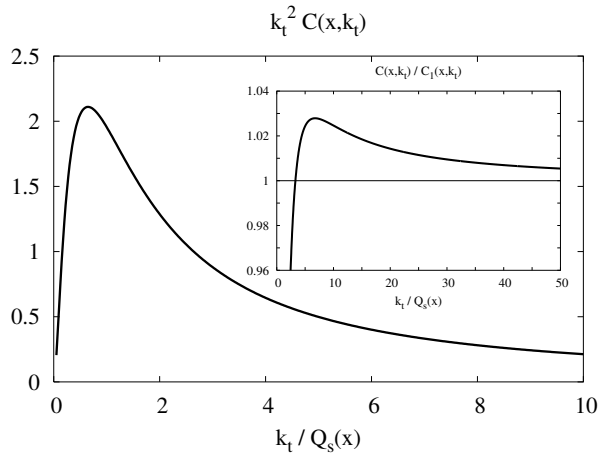}}
    \end{center} \caption{\label{fig:C-sat} The function $k_\perp^2
    C(x,k_\perp)$ in the model defined by eq.~(\ref{eq:edmond-mu2})
    (plotted against $k_\perp/Q_{s,_A}(x)$ thanks to its scaling
    properties). The insert in the upper right corner represents the
    ratio $C(x,k_\perp)/C_1(x,k_\perp)$.}
\end{figure}
That higher twists tend to push the spectrum towards higher momenta is
clearly seen in figure \ref{fig:C-sat}, where we have both represented
the function $k_\perp^2 C(x,k_\perp)$ (in the main plot), and the
ratio $C(x,k_\perp)/C_1(x,k_\perp)$ (in the smaller insert). However,
as the plot of $C(x,k_\perp)/C_1(x,k_\perp)$ also shows, the maximum
over the single scattering term is now of only $3$\%, compared to
$75$\% in the MV model (see the figure \ref{fig:Ca}).

Assuming that the nuclear enhancement ratio $R_{pA}$ is well
approximated by eq.~(\ref{eq:R-approx}), we can obtain its asymptotic
value at high $k_\perp$ from the leading term $C_1(x,k_\perp)$:
\begin{equation}
 R_{pA}\empile{\approx}\over{k_\perp\to \infty} 
A^{-1/3} \left(
\frac{Q_{s,_A}^2}{Q_{s,p}^2}
\right)^\gamma=
A^{(\gamma-1)/3}\; .
\label{eq:Rpa-asympt}
\end{equation}
Since $\gamma<1$, this asymptotic value is smaller than $1$. For
$\gamma\approx 0.64$ and $A\approx 6$, we expect $R_{pA}\approx 0.52$
at large momentum (this ratio would probably go to $1$ asymptotically
if we let the anomalous dimension $\gamma$ depend on transverse
momentum and go to $1$ at large momentum). Moreover, the ratio
$R_{pA}$ in this approximation is the ratio of two functions similar
to the one plotted in the upper right corner of figure
\ref{fig:C-sat}, rescaled horizontally by their respective $Q_s$ and
vertically by $Q_s^{2\gamma}$. Therefore, we expect the ratio $R_{pA}$
to be below its asymptotic value $A^{(\gamma-1)/3}$ at small
$k_\perp$, to eventually become larger than $A^{(\gamma-1)/3}$ and
reach a very mild maximum (of a few percent), and finally to reach the
asymptotic value from above.

\begin{figure}[htbp]
\begin{center}
\resizebox*{!}{6cm}{\includegraphics{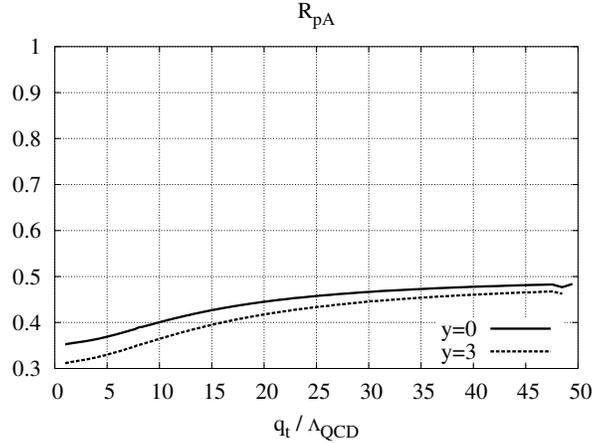}}
\end{center}
\caption{\label{fig:rpa-bfkl} The ratio $R_{pA}$ in the non-local
Gaussian model of eq.~(\ref{eq:edmond-mu2}).}
\end{figure}

We have studied the ratio $R_{pA}$ numerically, and
the results are displayed in the figure \ref{fig:rpa-bfkl}.
The asymptotic value of $R_{pA}$ is in relatively good agreement with
the crude estimate given in eq.~(\ref{eq:Rpa-asympt}). It seems also
that the very mild maximum we predicted on the basis of the
approximate formula of eq.~(\ref{eq:R-approx}) has been washed out by
the convolution present in the exact formula. The shape of the curve
now agrees qualitatively with BRAHMS data for $y=3$. However, it also
gives too much of a suppression at $y=0$!

\subsection{Interpretation of results}
We can summarize the key results of the model calculations in this
section as follows.

\begin{itemize}
  
\item Tree level partonic re-scattering responsible for the classical
Cronin effect in high energy deuteron-Gold collisions is computable in
the color glass condensate approach and when the weight for the color
sources is Gaussian, the formalism reduces to the familiar Glauber
formalism of independent multiple scatterings. In the simple
McLerran-Venugopalan model, a peak appears at $k_\perp\sim Q_{s,_A}$,
which is more pronounced for more central collisions.
  
\item Quantum corrections (due to small-$x$ evolution) to the tree
level partonic re-scattering can be ``naively'' included by letting
$Q_s\rightarrow Q_s(x)$. The parameterization of $Q_s(x)$ as a
function of $x$ that we have used is of the form first suggested by
Golec-Biernat and W\"usthoff but with the parameters from a recent fit
by Iancu, Itakura and Munier to the HERA data.  We call this model of
quantum evolution naive because it does not consider the possibility
that quantum evolution will change the form of the weight functional
from the Gaussian form of the MV model. This naive quantum evolution
model predicts a larger Cronin effect at larger rapidities (because
$Q_s$ grows as $x$ decreases) and is in sharp disagreement with the
RHIC d-Au data at forward rapidities~\cite{Debbe1}. 
  
\item Proper quantum evolution in the CGC is described by the JIMWLK
evolution equation. For the two-point correlator, of interest in
Cronin studies, the large $N$ limit is described by the
Balitsky-Kovchegov equation.  Like any other evolution equation, these
equations evolve from an initial condition at some large $x=x_0$. At
very small $x\ll x_0$, the system loses memory of its initial
conditions. The JIMWLK equation, in this limit, is well described by a
mean field approximation, and analytic expressions have been
derived. Interestingly, the weight functional in this extreme limit is
also a Gaussian, but with a non-local kernel. That is, the variance
$\mu_{_{A}}^2$ of the Gaussian sources is now a function of $k_\perp$:
$\mu_{_{A}}^2(x)\rightarrow \mu_{_{A}}^2(x,k_\perp)$. At low
$k_\perp$, $\mu_{_{A}}^2 \propto k_\perp^2 \ln (1+
(Q_s^2/k_\perp^2)^\gamma)$, where $\gamma$ is an anomalous dimension
($\gamma\approx 0.64$ for BFKL evolution). Because the weight is
Gaussian, the physics remains that of multiple independent
scatterings. However, as our calculation in section
\ref{sec:non-local} shows, the effects of multiple scatterings depend
very much on the shape of the spectrum. It follows that the
qualitative behavior of the ratio $R_{pA}$ is different from what it
is in the MV model with a local weight.  If we assume that $y=0$ and
$y=3$ at RHIC lie in this ``super saturated'' regime, we find a
suppression both at $y=0$ and $y=3$.  Thus while our results for $y=3$
are qualitatively close to the BRAHMS data, the $y=0$ results are
decidedly not close to the corresponding $y=0$ data.  Our result
therefore suggests that quantum evolution is not yet significant at
$y=0$. This of course was also evident from the disagreement of the
prediction of Kharzeev, Levin and McLerran with the RHIC d-Au data at
$y=0$. Our result merely provides additional insight into why one has
this disagreement.

\end{itemize}

From the comparison of the results of our simple studies with the
qualitative features of the RHIC d-Au data emerges the following
picture, which will have to be substantiated by more detailed (read
quantitative) computations and a thorough confrontation to the
experimental data. That the MV model works reasonably well at $x\sim
10^{-2}$ -- central rapidities at RHIC -- and reproduces key
qualitative features of central d-Au collisions suggests that it has
the right physics built in and that it could be a good model of the
initial conditions for quantum evolution as a function of rapidity.
The fact that at the same time the MV model fails badly at forward
rapidities is suggestive of important (and non-trivial) quantum
evolution effects that cannot be accounted for simply by a rescaling
of the saturation momentum. This interpretation is consistent with the
fact that the asymptotic mean-field solution of the JIMWLK evolution
equation is in qualitative agreement with the d-Au data at $y=3$.  In
particular, a prediction of this mean-field solution is that the ratio
$R_{pA}$ tends to $A^{(\gamma-1)/3}$ at large transverse momentum,
where $\gamma$ is the BFKL anomalous dimension.  If there is some
truth to this interpretation, one would also predict that the
suppression of the ratio $R_{pA}$ would change very little if even
larger rapidities were available (since $y=3$ seems to be already in
the asymptotic regime). This corroborates the numerical results
obtained previously from solutions of the BK equation by Albacete et
al~\cite{AlbacAKSW1}, and is also confirmed by Iancu, Itakura and
Triantafyllopoulos \cite{IancuIT1}. The preliminary RHIC d-Au data
appear to similarly show a rapid change at lower rapidities but a
weaker one at the higher rapidities probed.

\section{Summary and Outlook}
\label{sec:conclusions}
We have performed in this paper a systematic study of gluon production
in pA collisions within the framework of the Color Glass Condensate.
Working in the covariant gauge, we have solved the Yang-Mills
equations to lowest order in the dilute proton source but to all
orders in the nuclear source. Our explicit results for the gauge
fields are useful in computing quark production in pA collisions. Our
results on these are reported on in a companion paper. We have
computed gluon production and demonstrated explicitly how
$k_\perp$-factorization arises in this approach. We have demonstrated
explicitly the equivalence of calculations of gluon production in
different gauges.

The $k_\perp$-dependent distribution that describes the nucleus is
proportional to a two-point correlator of Wilson lines. This suggests
how small-$x$ quantum evolution effects can be taken into account in
studying proton-nucleus collisions in the CGC framework since the two
point correlators satisfy non-linear evolution equations with $x$, or
equivalently the rapidity $y$. We next studied in some detail the
Cronin effect in the CGC framework. A rapid transition from the regime
where multiple scattering effects are dominant to one where quantum
evolution effects are dominant is suggested by simple analytical
studies corroborating previous numerical studies.  If confirmed by
additional experimental results, this rapid transition has important
ramifications in the near future for heavy ion experiments at the LHC
and even for the proposed electron-heavy ion deeply inelastic
scattering experiments (eRHIC) proposed at BNL.

\section*{Acknowledgements}
R.~V.'s research was supported by DOE Contract No. DE-AC02-98CH10886.
We would like to thank I.~Balitsky, E.~Iancu, K.~Itakura,
J.~Jalilian-Marian, K.~Kajantie, Yu.~Kovchegov, T.~Lappi, S.~Munier,
D.~Triantafyllopoulos, and K.~Tuchin for useful discussions on this
and related topics.

\appendix

\section{Proof of equation (\ref{eq:ret-evol})}
\label{sec:green}
Let's assume we want to solve the equation:
\begin{equation}
(\square_x+iF(x)\partial_x^+) A(x)=J(x)\; ,
\label{eq:eq-AJ}
\end{equation}
with a source $J(x)$. The function $F(x)$ is some Hermitian
matrix-valued object, and we assume that it does not depend on $x^-$,
i.e. that it commutes with the derivative $\partial_x^+$.  We want to
find the function $A(x)$ for times $x^+>0$, given the value of
$A(x^+=0,x^-,\x_\perp)$ (and maybe of its first derivatives, since the
equation is of second order in derivatives) and the value of $A(x)$
and its first derivatives on the boundary of the ``spatial''
directions (here $x^-$ and $\x_\perp$). In order to do that, we must
start with the equations for $A(x)$ and for the free retarded
propagator\footnote{The equation for the propagator when one acts on
the right has been written by analogy with what happens when the
``external field'' is constant: in this case, the propagator is
invariant by translation and depends only on $x-y$. Hence, the first
derivatives with respect to $y$ enter with the opposite sign compared
to the first derivatives with respect to $x$, while the second
derivatives are the same.}:
\begin{eqnarray}
&&(\stackrel{\rightarrow}{\square}_y
+iF(y)\stackrel{\rightarrow}{\partial}\!\!{}_y^+) A(y)=J(y)\; ,\nonumber\\
&&G{}_{_{R}}(x,y)(\stackrel{\leftarrow}{\square}_y-iF(y)\stackrel{\leftarrow}{\partial}\!\!{}_y^+)=\delta(x-y)\; .
\end{eqnarray}
We have written the equation for the propagator with respect to its
second point, which imposes to put the term $F(x)$ on the right
because of its matrix nature. The arrows on the differential operators
indicate on which side the derivatives act.  Then, multiply the first
equation by $G{}_{_{R}}(x,y)$ (on the left), the second equation by
$A(y)$ (on the right), subtract them, and integrate over $y$ (with
$y^+$ starting at $0$). This gives:
\begin{eqnarray}
A(x)&=&\int\limits_{y^+>0}d^4y\; G{}_{_{R}}(x,y)J(y)\nonumber\\
&+&
\int\limits_{y^+>0}d^4y\big\{
G{}_{_{R}}(x,y)
(\stackrel{\leftarrow}{\square}_y
-iF(y)\stackrel{\leftarrow}{\partial}\!\!{}_y^+)
A(y)
\nonumber\\
&&\qquad\qquad
-
G{}_{_{R}}(x,y)
(\stackrel{\rightarrow}{\square}_y
+iF(y)\stackrel{\rightarrow}{\partial}\!\!{}_y^+)
 A(y)
\big\}\; .
\end{eqnarray}
This is a generalization to the case of eq.~(\ref{eq:eq-AJ}) of the
standard Green's theorem of electrostatics. The first term is the one
we would naively expect if the initial and boundary conditions for the
field and its derivative are zero. In order to deal with the second
term, we first write
$\square_y=2\partial_y^+\partial_y^--{\bs\nabla}_{y_\perp}^2$.  For
the term involving the transverse derivatives, we can write:
\begin{eqnarray}
&&G{}_{_{R}}(x,y)\stackrel{\rightarrow}{\bs\nabla}_{y_\perp}^2 A(y)
-
G{}_{_{R}}(x,y) 
\stackrel{\leftarrow}{\bs\nabla}_{y_\perp}^2
A(y)
\nonumber\\
&&\qquad\qquad=
\stackrel{\rightarrow}{\bs\nabla}_{y_\perp}\cdot\left[
G{}_{_{R}}(x,y)\stackrel{\rightarrow}{\bs\nabla}_{y_\perp} A(y)
- G{}_{_{R}}(x,y)
\stackrel{\leftarrow}{\bs\nabla}_{y_\perp}
A(y)
\right]\; .
\end{eqnarray}
We then perform the integration over $\y_\perp$ using Stokes theorem,
and the corresponding term becomes:
\begin{equation}
\int\limits_{y^+>0}dy^+dy^-
\oint\limits_{\partial{\mathbbm R}^2} d{\bs n}_\perp \cdot\left[
G{}_{_{R}}(x,y)\stackrel{\rightarrow}{\bs\nabla}_{y_\perp} A(y)
-G{}_{_{R}}(x,y)
\stackrel{\leftarrow}{\bs\nabla}_{y_\perp}
A(y)
\right]\; ,
\end{equation}
where $\partial{\mathbbm R}^2$ is the boundary in the direction of
transverse coordinates, and ${\bs n}_\perp$ a unit normal vector on
this boundary. Similarly, we can deal with the term in
$\partial_y^+\partial_y^-$ by first writing:
\begin{eqnarray}
&&G{}_{_{R}}(x,y) 
\stackrel{\leftarrow}{\partial}\!\!{}_y^+\stackrel{\leftarrow}{\partial}\!\!{}_y^-A(y)
-G{}_{_{R}}(x,y)
\stackrel{\rightarrow}{\partial}\!\!{}_y^+\stackrel{\rightarrow}{\partial}\!\!{}_y^- A(y)
\nonumber\\
&&\qquad\qquad
=\stackrel{\rightarrow}{\partial}\!\!{}_y^+( G{}_{_{R}}(x,y)
\stackrel{\leftarrow}{\partial}\!\!{}_y^-
A(y)
)
-\stackrel{\rightarrow}{\partial}\!\!{}_y^-(G{}_{_{R}}(x,y)\stackrel{\rightarrow}{\partial}\!\!{}_y^+ A(y))\; .
\end{eqnarray}
Therefore, the corresponding term gives:
\begin{eqnarray}
&&2\int\limits_{y^+>0}dy^+d^2\y_\perp\;
\left[
G{}_{_{R}}(x,y)
\stackrel{\leftarrow}{\partial}\!\!{}_y^- 
A(y)
\right]_{y^-=-\infty}^{y^-=+\infty}
\nonumber\\
&&
-2\int dy^-d^2\y_\perp\;
\left[G{}_{_{R}}(x,y)\stackrel{\rightarrow}{\partial}\!\!{}_y^+ A(y)\right]_{y^+=0}^{y^+=+\infty}\; .
\end{eqnarray}
Using the fact that $G{}_{_{R}}$ is a retarded propagator, the
boundary term at $y^+=+\infty$ is zero. Finally, we need to deal with
the term in $F(y)\partial_y^+$. We have:
\begin{eqnarray}
&&-i\int\limits_{y^+>0}d^4y\big\{
G{}_{_{R}}(x,y)F(y)\stackrel{\leftarrow}{\partial}\!\!{}_y^+A(y)
+
G{}_{_{R}}(x,y)F(y)\stackrel{\rightarrow}{\partial}\!\!{}_y^+A(y)
\big\}\nonumber\\
&&\qquad=
-i\int\limits_{y^+>0}dy^+ d^2\y_\perp
\left[G{}_{_{R}}(x,y)F(y)A(y)\right]_{y^-=-\infty}^{y^-=+\infty}\; .
\end{eqnarray}
Here, we have used the fact that $F(y)$ commutes with $\partial_y^+$
and the fact that $G_{_{R}}(x,y)$ is a retarded propagator.  If we
combine all the pieces together, we get:
\begin{eqnarray}
A(x)&=&\int\limits_{y^+>0}d^4y\; G{}_{_{R}}(x,y)J(y)\nonumber\\
&+&\int\limits_{y^+=0} 
dy^-d^2\y_\perp\;
G{}_{_{R}}(x,y)2\stackrel{\rightarrow}{\partial}\!\!{}_y^+ A(y)
\nonumber\\
&+&\int\limits_{y^+>0}dy^+dy^-
\oint\limits_{\partial{\mathbbm R}^2} d{\bs n}_\perp \cdot\left[
G{}_{_{R}}(x,y)\stackrel{\rightarrow}{\bs\nabla}_{y_\perp} A(y)
-G{}_{_{R}}(x,y)
\stackrel{\leftarrow}{\bs\nabla}_{y_\perp} A(y)
\right]\nonumber\\
&+&
\int\limits_{y^+>0}dy^+d^2\y_\perp\;
\left[
G{}_{_{R}}(x,y)(
2\stackrel{\leftarrow}{\partial}\!\!{}_y^- 
-iF(y)
)
A(y)
\right]_{y^-=-\infty}^{y^-=+\infty}\; .
\end{eqnarray}
If we assume that the field and its derivatives vanish fast enough at
infinity in the spatial directions, we get the simpler formula:
\begin{eqnarray}
A(x)&=&\int\limits_{y^+>0}d^4y\; G{}_{_{R}}(x,y)J(y)\nonumber\\
&+&\int\limits_{y^+=0} 
dy^-d^2\y_\perp\;G{}_{_{R}}(x,y)2\stackrel{\rightarrow}{\partial}\!\!{}_y^+ A(y)\; .
\label{eq:A-solution}
\end{eqnarray}
Eq.~(\ref{eq:ret-evol}) is a special case of this formula, where the
source term $J$ is zero. Naturally, in the above derivation, the
hyperplane $y^+=0$ could have been replaced by any plane of constant
$y^+$ on which we decide to set the initial condition.

\section{Simplification of equation (\ref{eq:Ai-1})}
\label{sec:B}
In order to simplify eq.~(\ref{eq:Ai-1}), we need to compute the
following convolution product:
\begin{eqnarray}
U\otimes (A_{_{A}}^-\cdot T)\otimes V
\equiv\int\limits_0^{\epsilon}dx^+
U(\epsilon,x^+)
(A_{_{A}}^-(x^+)\cdot T)
V(x^+,0)\; .
\end{eqnarray}
We have not written the transverse coordinates in this expression;
they are the same for the three factors and are irrelevant for the
following discussion. Let us first define
\begin{equation}
W(b^+,a^+|x^+)\equiv {\cal P}_+\left[\exp{i\frac{g}{2}}
\int\limits_{a^+}^{b^+}dz^+
(1+\theta(z^+-x^+))A_{_{A}}^-(z^+)\cdot T\right]\; .
\end{equation}
This object interpolates between $U(b^+,a^+)$ and $V(b^+,a^+)$:
\begin{eqnarray}
&&W(b^+,a^+|x^+)=V(b^+,a^+)\quad{\rm if\ }x^+\ge b^+\; ,\nonumber\\
&&W(b^+,a^+|x^+)=U(b^+,a^+)\quad{\rm if\ }x^+\le a^+\; .
\end{eqnarray}
We then have
\begin{eqnarray}
\frac{\partial W(b^+,a^+|x^+)}{\partial x^+}
&=&
-i\frac{g}{2}
W(b^+,x^+|x^+)
(A_{_{A}}^-(x^+)\cdot T)
W(x^+,a^+|x^+)
\nonumber\\
&=&
-i\frac{g}{2}
U(b^+,x^+)
(A_{_{A}}^-(x^+)\cdot T)
V(x^+,a^+)\; .
\end{eqnarray}
Then, integrating over $x^+$ from $0$ to $\epsilon$, we get:
\begin{eqnarray}
i\frac{g}{2}
[U\otimes (A_{_{A}}^-\cdot T)\otimes V]
=U(\epsilon,0)-V(\epsilon,0)\; .
\label{eq:UV-conv}
\end{eqnarray}

Let us now go back to eq.~(\ref{eq:Ai-1}). Using the fact that $0\le
v^+\le \epsilon$, we can approximate $\exp(iq^-x^+)\approx 1$. Using
eq.~(\ref{eq:UV-conv}), we have at this point:
\begin{eqnarray}
q^2 { A}^i(q)
&=&
2\int \!\!dv^-d^2\v_\perp\, e^{i(q^+v^--\q_\perp\cdot v_\perp)}
(V(\v_\perp)\!-\!U(\v_\perp))
\partial_v^i A_{p}^+(v)\nonumber\\
&+&
ig\int \!\!dv^-d^2\v_\perp
\,e^{i(q^+v^--\q_\perp\cdot v_\perp)}
\big\{
[U\otimes(\partial_v^i A_{_{A}}^-\cdot T)\otimes V](\v_\perp)
\nonumber\\
&&\qquad\qquad
-
[U\otimes(A_{_{A}}^-\cdot T)\otimes(\partial_v^i V)](\v_\perp)
\big\}A_{p}^+(v)\; .
\end{eqnarray}
The next step is to make more explicit the derivative $\partial_v^i
V$:
\begin{equation}
\partial_v^i V
=i\frac{g}{2}[V\otimes(\partial_v^i A_{_{A}}^-\cdot T)
\otimes V]\; ,
\end{equation}
which leads to\footnote{We use the fact that the convolution product
  defined above is associative in the following sense:
\begin{equation}
[F\otimes A\otimes[G\otimes B\otimes H]]=
[[F\otimes A\otimes G]\otimes B\otimes H]\; .
\end{equation}
This can be proven simply by permuting the integrals over the two
intermediate coordinates:
\begin{equation}
\int_0^\epsilon dx^+ \int_0^{x^+} dy^+=
\int_0^\epsilon dy^+\int_{y^+}^\epsilon dx^+\; .
\end{equation}
 }:
\begin{eqnarray}
&&[U\otimes(\partial_v^i A_{_{A}}^-\cdot T)\otimes V]
-[U\otimes(A_{_{A}}^-\cdot T)\otimes (\partial_v^i V)]
\nonumber\\
&&=
[U\otimes(\partial_v^i A_{_{A}}^-\cdot T)\otimes V]
-i\frac{g}{2}[U\otimes(A_{_{A}}^-\cdot T)\otimes
V\otimes(\partial_v^i A_{_{A}}^-\cdot T)\otimes V]
\nonumber\\
&&=i\frac{g}{2}
[V\otimes (\partial_v^i A_{_{A}}^-\cdot T)\otimes V]
=\frac{2}{ig}\partial_v^i V\; .
\end{eqnarray}

\section{Glauber interpretation of $C(k_\perp)$}
\label{sec:glauber}
It is fairly easy to write the Fourier transform
$\delta_{ac}C(p_\perp)$ of
$\big<U^\dagger(0)U(\x_\perp)\big>$ in a form that makes its
interpretation more intuitive in terms of the Glauber picture. Our
starting point is eq.~(\ref{eq:UUdagger}) (which is the most general
form one can obtain in the case of a Gaussian distribution of
$\rho_{_{A}}$):
\begin{equation}
C(p_\perp)=\int d^2\x_\perp
e^{i\p_\perp\cdot\x_\perp}
\exp\left[
-\frac{g^4N}{2\pi}
\int\limits_0^{+\infty}\frac{dk_\perp}{k_\perp^3}(1-J_0(k_\perp x_\perp))
\mu_{_{A}}^2(x,k_\perp)
\right]\; .
\end{equation}
At this point, it is convenient to factor out of $\mu_{_{A}}^2$ a constant
$\mu_0^2$ that sets the scale:
\begin{equation}
\mu_{_{A}}^2(x,k_\perp)\equiv \mu_0^2 f(x,k_\perp)\; .
\end{equation}
Typically, $\mu_0^2$ is given by $\mu_0^2=\rho L$, where $\rho$ is the
density in the target and $L$ is the longitudinal size of the target.
Then, we can introduce the following shorthands:
\begin{eqnarray}
&&
\sigma_{\rm tot}
\equiv\frac{g^4N}{2\pi}
\int\limits_0^{+\infty}\frac{dk_\perp}{k_\perp^3}f(x,k_\perp)\; ,
\nonumber\\
&&{ \sigma}(\x_\perp)
\equiv
\frac{g^4N}{2\pi}
\int\limits_0^{+\infty}\frac{dk_\perp}{k_\perp^3}
J_0(k_\perp x_\perp) f(x,k_\perp)\; .
\end{eqnarray}
$\sigma_{\rm tot}$ is the total cross-section of an incoming parton on
a parton of the target, and ${ \sigma}(\x_\perp)$ is the Fourier
transform of the differential cross-section (with respect to the
transverse momentum transfer). We can therefore write:
\begin{eqnarray}
C(p_\perp)&=&e^{-\mu_0^2 \sigma_{\rm tot}}
\int d^2\x_\perp e^{i\p_\perp\cdot\x_\perp}
e^{\mu_0^2 { \sigma}(\x_\perp)}\nonumber\\
&=&e^{-\mu_0^2 \sigma_{\rm tot}}
\int d^2\x_\perp e^{i\p_\perp\cdot\x_\perp}
\sum_{n=0}^{+\infty}\frac{\mu_0^{2n}}{n!}
\left[{ \sigma}(\x_\perp)\right]^n\; .
\end{eqnarray}
If we now replace each factor of ${ \sigma}(\x_\perp)$ by its
explicit expression as a Fourier transform of $\sigma(\k_\perp)\equiv
g^4 N f(x,k_\perp)/k_\perp^4$, we get:
\begin{eqnarray}
&&C(p_\perp)=e^{-\mu_0^2 \sigma_{\rm tot}}
\sum_{n=0}^{+\infty}\frac{\mu_0^{2n}}{n!}
\int
\frac{d^2\k_{1\perp}}{(2\pi)^2}\cdots\frac{d^2\k_{n\perp}}{(2\pi)^2}
\nonumber\\
&&\qquad\qquad\quad
\times(2\pi)^2\delta(\k_{1\perp}+\cdots+\k_{n\perp}-\p_\perp)
\sigma(\k_{1\perp})\cdots\sigma(\k_{n\perp})\; .
\label{eq:glauber}
\end{eqnarray}
Finally, if we notice that:
\begin{equation}
\frac{\mu_0^{2n}}{n!}=\rho^n \int\limits_0^L dz_1
\int\limits_{z_1}^L dz_2\cdots
\int\limits_{z_{n-1}}^L dz_n\; ,
\end{equation}
we see that the object $C(p_\perp)$ can be interpreted as the Glauber
form of the cross-section of a parton undergoing multiple independent
scatterings in the target (the index $n$ in the sum is the number of
such scatterings, and the $z_i$ are the longitudinal coordinates of
the scattering centers). Note that this correspondence is certainly
only valid when the functional distribution of the source
$\rho_{_{A}}$ is a Gaussian distribution (having a distribution with
non-zero $\big<\rho_{_{A}}\rho_{_{A}}\rho_{_{A}}\big>$,
$\big<\rho_{_{A}}\rho_{_{A}}\rho_{_{A}}\rho_{_{A}}\big>\cdots,$
introduces correlations between the different collisions and the
Glauber picture of independent successive collisions breaks down).
The form given in eq.~(\ref{eq:glauber}) for $C(p_\perp)$ also makes
obvious the following sum-rule:
\begin{equation}
\int\frac{d^2\p_\perp}{(2\pi)^2} C(p_\perp)=1\; .
\end{equation}
Indeed, integrating eq.~(\ref{eq:glauber}) simply removes the delta
function that was constraining the variables $\k_{i\perp}$, so that
integrating over each $\k_{i\perp}$ now gives $\sigma_{\rm tot}$. One
can then re-exponentiate the sum over $n$, which gives a factor
$\exp(\mu_0^2\sigma_{\rm tot})$ that compensates the prefactor
$\exp(-\mu_0^2\sigma_{\rm tot})$.

\end{document}